\documentclass[aps,prd,superscriptaddress,twocolumn]{revtex4-1}
\usepackage{graphicx}
\usepackage{amsmath}
\usepackage{amssymb}
\usepackage[usenames,dvipsnames]{xcolor}
\usepackage[per-mode = fraction]{siunitx}
\usepackage{chemformula}
\usepackage{bm}

\newcommand*{\VL}{V_\mathrm{L}}
\newcommand*{\Er}{E_\mathrm{r}}
\newcommand*{\EF}{E_\mathrm{F}}

\newcommand*{\Nc}{N_\mathrm{c}}
\renewcommand*{\Re}{\mathrm{Re}}
\renewcommand*{\Im}{\mathrm{Im}}
\newcommand*{\thyb}{t_\mathrm{hyb}}
\newcommand*{\cE}{\mathcal{E}}
\newcommand*{\work}{\mathcal{W}}
\newcommand*{\cED}{\mathcal{E}_\mathrm{D}}
\newcommand*{\cEF}{\mathcal{E}_\mathrm{F}}
\newcommand*{\cEvac}{\mathcal{E}_{\mathrm{vac}}}
\newcommand*{\e}{E}

\definecolor{light-gray}{gray}{0.55}

\begin{document}

\title{Effect of resonant impurity scattering of carriers on Drude peak broadening \\ in uniaxially strained graphene}

\author{V.O. Shubnyi}
\affiliation{Bogolyubov Institute for Theoretical Physics, National Academy of Science of
Ukraine, 14-b Metrologichna Street, Kyiv, 03680, Ukraine}
\affiliation{Department of Physics, Taras Shevchenko National University of Kyiv,
6 Academician Glushkov Avenue, Kyiv, 03680, Ukraine}

\author{Y.V.~Skrypnyk}
\affiliation{Bogolyubov Institute for Theoretical Physics, National Academy of Science of
Ukraine, 14-b Metrologichna Street, Kyiv, 03680, Ukraine}

\author{S.G.~Sharapov}
%\email{sharapov@bitp.kiev.ua}
\affiliation{Bogolyubov Institute for Theoretical Physics, National Academy of Science of Ukraine, 14-b Metrologichna Street, Kyiv, 03680, Ukraine}
\affiliation{Centre for Advanced 2D Materials, National University of Singapore, 6 Science Drive 2, Singapore 117546}

\author{V.M.~Loktev}
\affiliation{Bogolyubov Institute for Theoretical Physics, National Academy of Science of Ukraine, 14-b Metrologichna Street, Kyiv, 03680, Ukraine}
\affiliation{National Technical University of Ukraine KPI, 37 Peremogy Ave.,  Kyiv UA-03056, Ukraine}

\date{\today }

\begin{abstract}
An explanation is proposed for the recently observed in optical spectra of monolayer graphene
giant increase in the Drude peak width under applied uniaxial strain.
We argue that the underlying mechanism of this increase can be based on resonant scattering of carriers from
inevitably present impurities such as adsorbed atoms that can be described by the  Fano-Anderson model.
We demonstrate that the often neglected scalar deformation potential plays the essential role in this process.
The conditions necessary for the maximum effect of the giant Drude peak broadening
are determined. It is stressed that the effect is strongly enhanced when the Fermi level gets closer to the Dirac point.
Our theoretical analysis provides  guidelines  for functionalizing  graphene samples
in a way that would allow to  modulate efficiently the Drude peak width by the applied strain.
\end{abstract}

%\pacs{73.22.Pr, 73.20.At}

%81.05.ue 	Graphene (for structure of graphene, see 61.48.Gh; for phonons in graphene, see 63.22.Rc; for thermal properties, see %65.80.Ck; for graphene films, see 68.65.Pq; for electronic transport, see 72.80.Vp; for electronic structure, see 73.22.Pr; for %optical properties, see 78.67.Wj)

%65.80.Ck 	Thermal properties of graphene

% 73.20.At 	Surface states, band structure, electron density of states

% 72.80.Vp 	Electronic transport in graphene

%\email{}

%\keywords{GRAPHENE}

\maketitle

\section{Introduction}

It is not surprising that after 14 years from the discovery of graphene there is a shift from the fundamental towards the applied research.
Despite the fact that the electronic subsystem of graphene is well-understood in the tight-binding approximation, some fundamental
issues of its physics still remain unresolved.
A curious mixture of fundamental and applied
physics that explores possibilities to use strain for controlling
physical properties of graphene was recently coined out as {\em straintronics} \cite{Vozmediano2010PR,Amorim2015PR,Naumis2017RPP,Wang2019}.

Carbon atoms in the monolayer graphene constitute a honeycomb lattice due
to sp$^2$ hybridization of their orbitals.
These in-plane hybridized orbitals form $\sigma$ bonds that are responsible for outstanding mechanical strength and stiffness of graphene, which is able to sustain elastic deformations in excess of 20\% \cite{Lee2008Science}.
The graphene's electronic, e.g., transport and optical (see the review in Ref.~\onlinecite{Basov2014RMP})
properties are among the most desired to control for both fundamental physics and technology.
These properties are governed by the electrons in the valence and conduction bands.
The latter are formed by the remaining 2p$_z$ orbitals (making $\pi$ bonds),
which are arranged perpendicular to the graphene sheet. Since there is no significant
mixing between states belonging to 2p$_z$ and 2sp${}^2$ bands, the electronic
properties in vicinity of the Fermi energy  can be well
described by a tight-binding model with only one orbital per atom. Then, the impact of the
deformation on the electronic properties can be modeled by taking into account
the strain dependence of the hopping parameters as well as on-site energies of these orbitals.

The uniaxial stretch is the simplest strain configuration for
theoretical and experimental study.
The optical conductivity of uniaxially strained graphene was studied
theoretically in
Refs.~\onlinecite{Pellegrino2010PRB,Pereira2010EPL,Pellegrino2011PRB,Pellegrino2011HP,Oliva2014JPCM}.
 The main result of these works
is the anisotropic renormalization of the interband conductivity as a
function of magnitude and direction of strain.
These simple enough predictions seem to be in agreement with the measurements
\cite{Ni2014AdvM}
of transparency in the visible range in large-area chemical vapor
deposited (CVD) monolayer
graphene pre-strained on a polyethylene terephthalate (PET) substrate.
The uniaxial strain of the order of $0.5\%$ results in a transparency
change of $0.1\%$.
%Notice that that the nominal strain imposed on a substrate differs
%from the actual strain of graphene.
%To overcome this difficulty the experimentalists use the Raman
%spectroscopy that
%provides an accurate probe of strain in graphene.

The production of monolayer 30-inch graphene films
\cite{Bae2010NatNano}
has opened a route to various practical applications.
However, CVD graphene exhibits low carrier mobility compared to
exfoliated graphene. This indicates that the carrier
scattering in the former case is more substantial.
A recent experimental study of the far-infrared transmission
spectra of large-area CVD monolayer graphene on a PET substrate
\cite{Kuzmenko20172DMat} revealed a new rather strong effect that has
not been theoretically expected.
It was found that the Drude-peak width increases by more than $10\%$
per $1\%$ of applied uniaxial strain, while the Drude
weight and, therefore, the Fermi energy remain unchanged.
To exclude the effect of relaxation of wrinkles and folds, directly
seen by atomic-force microscopy, the actual strain was measured using
the Raman spectroscopy.
Possible sources of electron scattering in graphene that may be
responsible for the observed
effect were theoretically analyzed in
Ref.~\onlinecite{Kuzmenko20172DMat}. They include short-range point defects,
long-range charged impurities, acoustic phonons,
surface phonons in the substrate and grain boundaries.
It was suggested that while the effect of surface phonons in
the substrate cannot be excluded, the dominant increase of the width
is due to
scattering from charged impurities related to the reduction of the
effective graphene-PET distance.

Another infrared spectroscopy study of carrier scattering of
large-area CVD monolayer graphene
on SiO$_2$/Si substrate was recently done in Ref.~\onlinecite{Yu2016PRB}. The
optical carrier scattering rate
as a function of the carrier density was studied, although its  strain
dependence was not investigated.

The purpose of this work is to put more emphasis on the role of short-range impurities in the
Drude-peak broadening.  For this task, the point defects
are assumed to be either chemically substituted carbon atoms, including their absence, i.e. vacancies, or
adsorbed atoms or molecules on a graphene sheet. They originate both as
by-products of a fabrication method and from exposure to an environment.

A significant amount of work done on the electron transport
in graphene had shown that it is strongly affected by the resonance impurity states
(see, e.g., Refs.~\onlinecite{Robinson2008PRL,Wehling2010PRL} and the reviews in
Refs.~\onlinecite{Peres2010RMP,Skrypnyk2018FNT}).
The corresponding partial scattering rate of carriers
is significantly enhanced in a vicinity of the impurity resonance energy
\cite{Skrypnyk2006PRB,Skrypnyk2007PRB,Skrypnyk2007FNT}. At that, the resonance energy itself
strongly depends on the behavior of the unperturbed density of states (DOS).
Thus, we expect that even  small variation of the DOS caused by strain may
result in a large change of the Drude-peak width.

The paper is organized as follows.
In Sec.~\ref{sec:model} we present model Hamiltonians that correspond to graphene with point defects.
In Sec.~\ref{sec:strain} we describe modification of the basic parameters of the model under mechanical strain, and discuss
its main implications.
In Sec.~\ref{sec:formalism} we discuss the formalism employed for the impurity effect study
and present analytical expressions for the main quantities that we consider in this paper.
The results are provided in Sec.~\ref{sec:results}.
In Sec.~\ref{subsec:results-lifshitz}, we present results for  the Lifshitz model.
In particular, we discuss the Born approximation as a weak scattering limit.
In  Sec.~\ref{subsec:results-fano}, we analyse the Fano-Anderson
model. In Conclusions, Sec.~\ref{sec:concl}, we give a concise summary of the obtained results.

\section{Models}
\label{sec:model}
We start with the Hamiltonian
\begin{equation}
\mathrm{H} = \mathrm{H}_{\mathrm{host}} + \mathrm{H}_{\mathrm{imp}},
\end{equation}
where $\mathrm{H}_{\mathrm{host}}$ is the  Hamiltonian for
electrons in strained graphene
and $\mathrm{H}_{\mathrm{imp}}$ is the impurity Hamiltonian.
In turn, the Hamiltonian $H_{\mathrm{host}}$ of strained
graphene consists of the two terms,
\begin{equation}
\label{eqn:hamiltonian_host}
\mathrm{H}_{\mathrm{host}} = \mathrm{H}_{\mathrm{hop}} + \mathrm{H}_{\mathrm{pot}}.
\end{equation}
The conventional tight-binding Hamiltonian $\mathrm{H}_{\mathrm{hop}}$ for $\pi$ orbitals of carbon
reads as
\begin{equation} \label{eqn:hamiltonian_0}
\mathrm{H}_{\mathrm{hop}}   = -\sum_{\langle i,j \rangle} t_{ij}^{\phantom\dagger} (\hat{c}_{A,
i}^\dagger \hat{c}_{B,j}^{\phantom\dagger} + \hat{c}_{B,j}^\dagger \hat{c}_{A,i}^{\phantom\dagger}
),
\end{equation}
where $i,j$ run over $N/2$ lattice cells,
indices $A$ and $B$ enumerate the sublattices, operator
$\hat{c}_{\alpha,i}^\dagger$ ($\hat{c}_{\alpha, i}$)
creates (annihilates) an electron at the corresponding lattice site,
%the spin index is omitted for brevity,
$\langle i,j \rangle$ denotes summing over nearest neighbors, $t_{ij}$ is the strain-dependent hopping amplitude.
Details on how deformation is included in the
Hamiltonian and its expansion in the vicinity of the Dirac point are presented in
Sec.~\ref{sec:strain} below.

The potential term is
\begin{equation} \label{eqn:hamiltonian_pot}
\mathrm H_{\mathrm{pot}}   =
\sum_{i, \alpha} U^{\phantom\dagger}_{\alpha i}
\hat{c}^\dagger_{\alpha i} \hat{c}^{\phantom\dagger}_{\alpha i},
\end{equation}
where $\alpha=A, B$
and $U_{\alpha i}$ is the on-site deformation-dependent potential.
$U_{\alpha i}$ consists of the strain-independent part $\cED$, which determines the energy of the Dirac point in unstrained graphene, and the strain-dependent part, which can be related to the interaction of the electrons with  long-wavelength acoustic phonons \cite{Suzuura2002PRB}.

In our description we assume that there is no mixing between the spin states, so the spin label can be omitted.
The corresponding two-fold degeneracy is taken into account when appropriate.

Graphene with point defects, in particular impurities that chemically
substitute the carbon atoms or vacancies,
can be modeled as a substitutional binary alloy with a diagonal
disorder. In this simple model, the hopping integrals for
the carbon-defect and defect-defect hoppings do not differ from the
clean graphene case.
The corresponding impurity Hamiltonian $H_{\mathrm{imp}}$ in this description, widely referred to as the Lifshitz model \cite{Lifshitz1964AP},  reads as
\begin{equation}
\label{eqn:Hamiltonian-Lifshitz}
\mathrm H_{\mathrm{L}} = \VL \sum_{l, \alpha} \eta^{\phantom\dagger}_{\alpha l}
\hat{c}^\dagger_{\alpha l} \hat{c}^{\phantom\dagger}_{\alpha l},
\end{equation}
where $l =1,\ldots ,N/2$ runs over the lattice cells with two atoms
per cell, $\VL$ is the impurity
potential, and $\eta_{\alpha l}$ is unity on the sites occupied by the
impurities and zero otherwise. In the model we use, the impurity
potential
$\VL$ is the same for every site occupied by an impurity.

The impurities are supposed to be distributed between lattice sites
without any correlation.
Accordingly,
%the value
$ \eta_{\alpha l}$ equals one with the
probability $c$
and zero with the probability $(1-c)$. Thus, the probability $c$
corresponds to the impurity concentration
per site and does not depend on the strain. For a large system with
$N$ sites, the total amount of impurities goes to $cN$.

%\subsection{Fano--Anderson model}
%\label{sec:Fano}

To describe adsorbed atoms or molecules on the graphene sheet, we employ the
Fano-Anderson impurity model \cite{Fano1961PRev}. It introduces a possibility
for an electron to transfer to an additional energy level that belongs to the
adsorbed impurity bound to a host atom. The impurity part of the Hamiltonian $H_{\mathrm{imp}}$
describing the adatoms and their interaction with the host reads as
\begin{equation}
\label{eqn:hamiltonian-fano}
H_{\mathrm{FA}} = \sum_{l, \alpha} \eta_{\alpha l}
\left[ U_{\alpha l}^\mathrm{imp} \,
d^\dagger_{\alpha l} d^{}_{\alpha l} + (t_\mathrm{hyb} d^\dag_{\alpha l} c^{}_{\alpha l}  + t_\mathrm{hyb}^\ast c^\dag_{\alpha l} d^{}_{\alpha l})\right].
\end{equation}
Here, index $l =1,\ldots ,N/2$ spans over the lattice cells, the
parameter $\eta_{\alpha l}$ is used in the same sense as in
Eq.~(\ref{eqn:Hamiltonian-Lifshitz}),
$t_\mathrm{hyb}$ is the hopping integral between the adatom and the host, $U_{\alpha l}^\mathrm{imp}$ is the potential on the adatom site,
$d^\dagger_{\alpha, l}$ and $d^{}_{\alpha l}$ are the creation and annihilation operators for this level.

\section{Uniaxially strained graphene}
\label{sec:strain}
Let us recapitulate the main results necessary for the description of uniaxially strained graphene and discuss a possible role played by the deformation potential term.
We choose the coordinate system so that the zigzag direction of the
honeycomb lattice is
parallel to the $Ox$ axis (similar to Refs.~\onlinecite{Vozmediano2010PR,Naumis2017RPP,
	Pellegrino2010PRB,Pereira2010EPL,
	Pellegrino2011PRB,Pereira2009PRB}).
The tensile stress, $\mathbf{T} = T \cos \theta \mathbf{e}_x +  T
\sin \theta \mathbf{e}_y $, can be applied at an arbitrary
angle  $\theta$ to the $Ox$ axis [$\mathbf{e}_{x,y}$ are the unit vectors in the
$Ox(y)$ directions]. In the following, we will also refer to the
principal coordinate
system $Ox^\theta y^\theta$ in which $\mathbf{T}$ is aligned along
the $Ox^\theta$ axis: $\mathbf{T} = T \mathbf{e}_x^\theta$.

Since we consider uniform planar strain, the components of two-dimensional strain tensor $\bar{ \bm{\varepsilon}}$
are position independent. Accordingly, the displacement vector
$\mathbf{u}(\mathbf{x})$ reads as
$\mathbf{u}(\mathbf{x}) = \bar{\bm{\varepsilon}} \cdot \mathbf{x}$.
Thus, the actual position of an atom
$\mathbf{x}^\prime =\mathbf{x} +\mathbf{u}(\mathbf{x})$ can be written as
$\mathbf{x}^\prime = (\bar{ \mathbf{I}} + \bar{\bm{\varepsilon}}) \cdot
\mathbf{x}$, where $\bar{ \mathbf{I}}$ is the unit
$2 \times 2$ tensor.

It turns out that the planar deformation of the hexagonal crystal in
the basal plane is determined by two
independent stiffness (compliance) tensor components; in other words,
it behaves as an isotropic planar solid \cite{Landau.book}.
Therefore, as we will find, the DOS is independent of the direction
of the applied strain.

In the principal coordinate system, $Ox^\theta y^\theta$, the only nonzero deformations are the longitudinal deformation, $\varepsilon_{xx}^\theta = T S_{xxxx}$, and Poisson's transverse contraction, $\varepsilon_{yy}^\theta = T S_{xxyy}$.
Here $S_{xxxx}$ and $S_{xxyy}$ are the two independent nonzero components
of the compliance tensor.
Accordingly, the strain tensor can be expressed in terms of the strain $\varepsilon \equiv \varepsilon_{xx}^\theta$ and
the Poisson's ratio, $\nu = - S_{xxyy}/S_{xxxx}$ as follows:
\begin{equation}
\bar{\bm{\varepsilon}}^\theta = \varepsilon \left(
                                            \begin{array}{cc}
                                              1 & 0 \\
                                              0 & -\nu \\
                                            \end{array}
                                          \right).
\end{equation}
A positive value of $\varepsilon$ means the strain is tensile, while a negative value corresponds to compressive strain.
In the original lattice coordinate system the strain tensor reads as
\begin{equation}
\bar{\bm{\varepsilon}} =
\varepsilon\left(
             \begin{array}{cc}
               \cos^2 \theta -\nu \sin^2 \theta & (1+\nu) \cos \theta \sin \theta \\
               (1+\nu) \cos \theta \sin \theta & \sin^2 \theta  - \nu \cos^2 \theta \\
             \end{array}
           \right).
\end{equation}

As pointed out in Ref.~\onlinecite{Pereira2009PRB}, when the stress is
induced on graphene by mechanically acting on the
substrate, the relation between strain and stress is given by the material parameters of the substrate, rather
than the intrinsic properties of graphene. The relevant tuning parameter in this case is the tensile strain,
$\varepsilon$. Furthermore, a small Poisson's ratio ($\nu=0.16$ in graphite \cite{Pereira2009PRB}) may possibly be
even smaller in graphene on substrate \cite{Kuzmenko20172DMat}.

As we stretch the sample, its area increases from $NS_0$ to $N S_\varepsilon$.
Here, $S_\varepsilon$ is the unit-cell area of uniformly strained
graphene,
which is related to the unit-cell area of the pristine graphene, $S_0
= 3 \sqrt{3} a^2/2$, by the relation
\begin{equation}
S_\varepsilon = S_0 (1 + \mbox{tr} \bar{\bm{\varepsilon}}) = S_0 [1+  (1-\nu) \varepsilon].
\end{equation}
Specific quantities per unit area, such as carrier density, impurity density, the density of states, etc., are affected
by this change, even though the total number of charges or impurities might have remain unchanged under the strain.
Thus, it is preferable to count such quantities per number of atoms. For example, we specified the quantity of impurities
by means of the impurity concentration $c$, i.e., a ratio of the number of the impure sites to the total number of sites.
Obviously, it does not change with the deformation, although the density of point defects per unit area
$n_{\mathrm{pd},\varepsilon} = c (2/S_\varepsilon)$ changes.

\subsection{Hopping Hamiltonian}

The Hamiltonian $H_\mathrm{hop}$ (\ref{eqn:hamiltonian_0}) contains the hopping integrals $t_{ij}$. In the absence of deformation,
$t_{ij}$ are independent of position of the neighbors. Thus, $|t_{ij}| = \mathrm{const} = t_0$ for every pair of nearest neighbors $\langle i, j \rangle$. The value $t_0 \approx \SI{2.7}{eV}$ is usually chosen \cite{Andrei2012RPR} to match the tight-binding band structure to results obtained from the first-principleі calculations.

The uniform uniaxial deformation preserves the translational symmetry, but breaks the rotational symmetry of the honeycomb lattice.
In the absence of deformation, the nearest-neighbor vectors that connect an $A$ atom to the three $B$ neighbors are
$\bm \delta_1^{0} =  a_0(0,1)$, $\bm \delta_2^{0} =  a_0 (\sqrt{3}/2,-1/2)$, and $\bm \delta_3^{0} =  a_0 ( -\sqrt{3}/2,-1/2)$,
where $a_0 = \SI{1.42}{\AA}$ is the distance between the nearest carbon atoms in undeformed graphene lattice. Due to
the uniform strain, the three nearest $B$ sites change their positions with respect to the $A$ sites,
so that the new vectors are $\bm \delta_n^\varepsilon = (\bar{ \bm{I}} + \bar{\bm{\varepsilon}}) \bm \delta_n^0$.
Whereas $|\bm \delta_1^{0}|=|\bm \delta_2^{0}|=|\bm \delta_3^{0}|=a_0$, the vectors
$\bm \delta_n^\varepsilon$ differ in their lengths in a general case.
Accordingly, there are three separate hopping integrals $t_n$ ($n=1,2,3$). Each one can be represented as a function of the distance between the neighbors, i.e., $|\bm \delta_n^\varepsilon|$. For a small deformation, we can use the first-order expansion in strain \cite{Naumis2017RPP}:
\begin{equation}
\label{hopping-n}
t_n   \approx
t_0 - \frac{\beta t_0}{a_0^2} \bm{\delta}_n^0 \cdot \bar{\bm{\varepsilon}} \cdot \bm{\delta}_n^0.
\end{equation}
Here, $\beta = (- \partial \ln t/\partial \ln |\bm \delta_n^\varepsilon|) |_{|\bm \delta_n^\varepsilon| = a_0}$ is the dimensionless Gr\"{u}neisen parameter. The values for this parameter vary between $2$ and $4$ across the literature; we will use $\beta = 3$ for our studies.

The effective  Hamiltonian describing uniformly strained graphene in the momentum representation can be obtained similarly
to the case of pristine graphene \cite{Amorim2015PR,Pellegrino2011PRB,Linnik2012JPCM,Oliva2013PRB,Naumis2017RPP}.
To derive a correct dispersion relation for electronic excitations near the Dirac points, one has to account for their shift in the $\mathbf{k}$ space \cite{Oliva2013PRB} from the initial positions $\mathbf{K}_{\pm}^0 = (\pm 4 \pi / (3 \sqrt{3}  a_0) , 0)$
in the undeformed graphene to the new positions at $\mathbf{K}_{\pm}^{\mathrm{D}} = (1 - \bar{\bm{\varepsilon}})\mathbf{K}_{\pm}^0 \pm \mathbf{A}^\mathrm{ps}$ \cite{Oliva2013PRB}. Here, $\mathbf{A}^\mathrm{ps} = (\beta /2 a_0) (\varepsilon_{xx} - \varepsilon_{yy},  -2 \varepsilon_{xy})$ is the strain-induced
vector potential \cite{Suzuura2002PRB}.
Thus, the momentum representation of the hopping Hamiltonian~(\ref{eqn:hamiltonian_0}) at $\mathbf{K}_{+}^{D}$ point reads as
\begin{equation}
\label{Dirac-strain}
H_\mathrm{hop}(\mathbf q; {\mathbf{K}_{+}^D})= \hbar v_0 \bm{\sigma} \cdot [1 - (\beta - 1)\bar{\bm{\varepsilon}}  ] \cdot \mathbf{q}, \quad |\mathbf{q}| a_0 \ll 1,
\end{equation}
where Pauli matrices $\bm{\sigma} = (\sigma_1, \sigma_2) $ act in
the sublattice space, $v_0 = 3 t_0 a_0 /( 2 \hbar)$ is the Fermi
velocity in the pristine graphene, and $\mathbf{q}$ is the
wave-vector measured from the shifted Dirac point
$\mathbf{K}_{+}^{D}$.
The corresponding Hamiltonian for $\mathbf{K}_{-}^D$ point can be
written by substituting $\bm{\sigma}  \to \bm{\sigma}^\ast =
(\sigma_1, -\sigma_2) $. The effect of strain in Eq.~(\ref{Dirac-strain}) is taken into account both via the $\beta$-independent term caused by the deformation of the unit cell of graphene lattice, and the $\beta$-dependent term caused by the changes in the hopping parameters (\ref{hopping-n}). Both contributions have the same order of magnitude.

In the absence of strain, the conical Dirac spectrum reads $\cE (\mathbf{q}) =  \pm \hbar v_0 |\mathbf q| + \cED$. Here $\cED$ is the strain-independent part of the potential $U_{\alpha l}$ defined below Eq.~(\ref{eqn:hamiltonian_pot}). The corresponding low-energy DOS per spin and unit area of unstrained graphene is $\rho_0(E) = |\cE - \cED|/(\pi \hbar^2 v_0^2)$.

For the uniaxially strained graphene, the spectrum distorts into elliptical cones,
$\cE(\mathbf{q}) = \pm \hbar \sqrt{v_\parallel^2 q_\parallel^2 + v_\perp^2 q_\perp^2} + \cED$, where
$q_\parallel \, \, (q_\perp)$ is the wave-vector component parallel (perpendicular)
to the direction of applied strain. The corresponding
components of the anisotropic Fermi velocity are $v_\parallel = v_0 [1 - (\beta-1)\varepsilon]$ and
$v_\perp = v_0 [1+ \nu (\beta-1) \varepsilon]$. \cite{Naumis2017RPP}

The only effect of the Dirac cone distortion on the DOS of the strained graphene (per spin and unit area), $\rho_\varepsilon (\cE)$, is the renormalization of its slope \cite{Naumis2017RPP,Oliva2014JPCM,Juan2013PRB,Shah2013MPLB}
\begin{equation}
\label{DOS-area}
\rho_\varepsilon (\cE)
% = 1 + (\beta-1) \mbox{tr} \bar{\bm{\varepsilon}}
= \frac{v_0^2}{ v_\parallel v_\perp} \rho_0 (\cE)
\approx [1 + (\beta -1) (1 - \nu) \varepsilon] \frac{|\cE -  \cED|}{\pi \hbar^2 v_0^2}.
\end{equation}
It is theoretically predicted that for large strain, $\varepsilon \geq 23 \% $, a gap in the quasiparticle spectrum opens \cite{Pereira2009PRB,Pellegrino2010PRB}. Here, we assume that the strain is small enough, and do not consider such qualitative changes in the spectrum.

In what follows, it is more convenient to consider the DOS per one atom:
\begin{equation}
\label{DOS-atom-cell}
D_\varepsilon (\cE) = \frac{S_\varepsilon }{2}  \rho_\varepsilon (\cE) = \frac{|\cE- \cED|}{W^2_\varepsilon}, \quad |\cE - \cED| \ll W_\varepsilon.
\end{equation}
The slope of this function, as follows from the normalization condition for the total number of states
$\int_{\cED- W_\varepsilon}^{\cED + W_\varepsilon} D_\varepsilon(\cE) d \cE = 1$,
equals to the reverse square of the strain-dependent effective bandwidth,
\begin{equation}
W_\varepsilon = W_0  \sqrt{1- \beta (1-\nu) \varepsilon } .
\label{eqn:w-epsilon}
\end{equation}
Here, $W_0 =(\sqrt{3} \pi )^{1/2} t_0  \simeq 2.33 t_0$ is
the effective bandwidth in the absence of strain.
It is essential that, in contrast to the Hamiltonian (\ref{Dirac-strain}) and the DOS per unit area (\ref{DOS-area}), the DOS (\ref{DOS-atom-cell}) does not depend on size or shape of the unit cell,
and depends only on the values of hopping integrals.

\begin{figure}[!h]
\raggedleft{
\includegraphics[width=1.\linewidth]{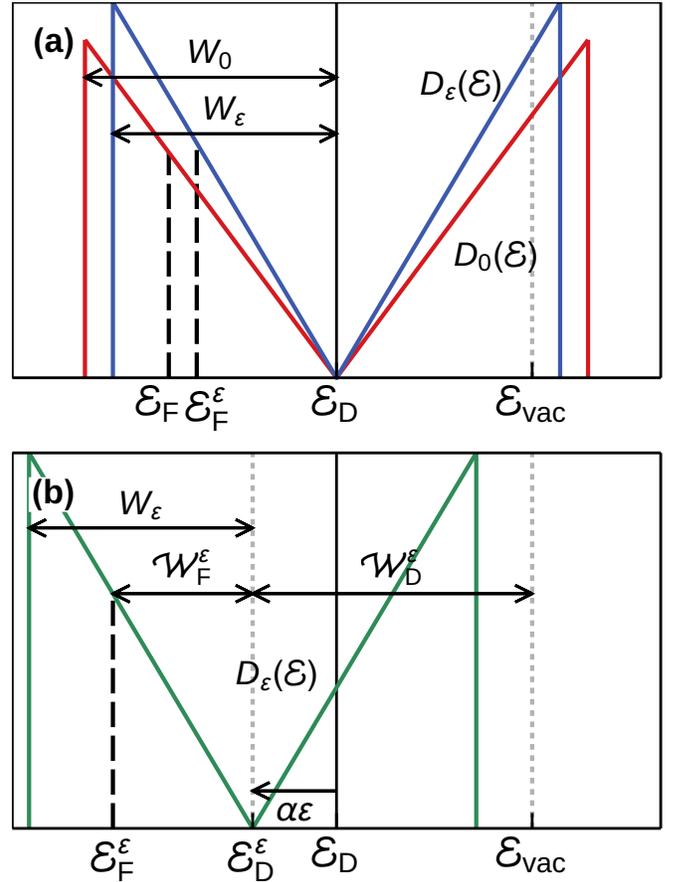}}
\caption{The influence of uniaxial strain %$\varepsilon >0$
on the DOS, $D_\varepsilon(E)$.
(a) Only the hopping term is taken into account.
The triangular DOS of unstrained (strained) graphene is shown by the red (blue) lines.
The unperturbed bandwidth is $W_0$ and the
position of the Fermi level $\cEF$ corresponds to the hole-doped sample. The tensile, $\varepsilon>0$,
strain results in the  decrease of effective bandwidth, $W_\varepsilon < W_0$.
Since the number of carriers is fixed, the position of the Fermi level
$\cEF^\varepsilon > \cEF$
shifts to the right in the hole-doped case.
(b) Both the hopping term and deformation potential  are included.
The latter results in the shift of the Dirac point from $\cED$ to the position $\cED^\varepsilon = \cED + \alpha \varepsilon$.
The position of the Fermi level shifts to the value
$\cEF^\varepsilon$ as described by Eq.~(\ref{eqn:efermi-epsilon}).
For a doped strained graphene the work function $\work = \work_D^\varepsilon + \work_F^\varepsilon $.
 }
\label{fig:1}
\end{figure}
In Fig.~\ref{fig:1} we illustrate how the strain modifies the DOS of clean graphene. In particular,
Fig.~\ref{fig:1}(a) shows only the effect of hopping integral modification, whereas Fig.~\ref{fig:1}(b)
adds the effect of the deformation potential. The latter will be discussed in Sec.~\ref{sec:deformation-potential}.

We note that the bandwidth $W_0 \approx \SI{6.3}{eV}$ for $t_0 = \SI{2.7}{eV}$ in the model with
the triangular DOS is noticeably less than the graphene bandwidth, which is
equal to $3t_0 \approx \SI{8.1}{eV}$ in the tight-binding approximation.
As it was discussed above, $\nu \leq 0.16$. However, in what follows
we will simply assume that $\nu=0$. If necessary,
actual value of $\nu$ can be easily restored by replacing $\varepsilon \to
\varepsilon (1 -\nu)$.

\subsection{Deformation potential}
\label{sec:deformation-potential}

In addition to the outcome of the hopping integral reduction, we have to take into account the effect
of strain on the potential term (\ref{eqn:hamiltonian_pot}) in the host Hamiltonian (\ref{eqn:hamiltonian_host}).
To fulfill this task, the conventional description by the deformation potential
(see, e.g., Ref.~\onlinecite{Suzuura2002PRB}) is employed here.
For uniaxial strain, the potential energy does not depend on the lattice site.
Thus, the Hamiltonian (\ref{eqn:hamiltonian_pot}) acquires the form
\begin{equation}
H_\mathrm{pot} = \sum_{i, \alpha} (\alpha \textrm{tr} \bar{\pmb{\varepsilon}} + \cED )
c^\dagger_{\alpha i} c^{}_{\alpha i}.
\label{eqn:deformation-potential}
\end{equation}
Here, $\alpha$ is the deformation potential.
In the momentum representation [cf. Eq.~(\ref{Dirac-strain})], $H_\mathrm{pot}$
can be expressed as
$(\alpha \varepsilon + \cED)$ multiplied by the $2\times 2$ unit matrix $\sigma_0$,
which is justified by the symmetry between sublattices. Accordingly, $H_\mathrm{pot}$
is also referred to as the scalar potential \cite{Vozmediano2010PR,Amorim2015PR}.

The presence of the uniform deformation potential due to the uniaxial strain shifts the whole distorted conical spectrum
along the energy axis, so that the Dirac point energy defined as $\cE(\mathbf{q} =0)$ moves from $\cED$ to
\begin{equation}
\cED^\varepsilon = \cED + \alpha \varepsilon,
\label{eqn:dirac-energy-shift}
\end{equation}
which can be clearly seen in Fig.~\ref{fig:1}(b).
Accordingly, the difference $\cE - \cED$  (see, e.g. the DOS (\ref{DOS-atom-cell}))
should be replaced by $\cE - \cED^\varepsilon$, where it applies.

Theoretical values of the deformation potential  $\alpha$ recited in the review \cite{Amorim2015PR}
are rather inconsistent between sources and vary in a fairly wide range from \SIrange{0}{20}{eV}.
One can estimate the value of $\alpha$ from the {\it ab initio} calculations \cite{Choi2010PRB}, which show that a $12\%$ uniaxial strain results in increase of the work function $\work_D^\varepsilon $ by $\SI{0.3}{eV}$.
In the undoped graphene, $\work_D^\varepsilon $  is defined as the difference
between the local vacuum energy level $\cEvac$ and the Dirac point energy $\cED^\varepsilon $, viz.
$\work_D^\varepsilon = \cEvac - \cED^\varepsilon$.
Thus, $\work_D^\varepsilon = \cEvac - \cED  - \alpha \varepsilon = \work_0 - \alpha \varepsilon$,
where $\work_0$ is the work function for undoped and unstrained
graphene.  Accordingly, the deformation potential $\alpha = - d \work_D^\varepsilon/d \varepsilon \approx
\SI{-2.5}{eV}$.

It was demonstrated in Ref.~\onlinecite{Samaddar2016Nanoscale} that the work function varies in one-to-one correspondence
to the position of the Fermi level in monolayer graphene.
This relation was verified down to the nanometer scale, where due to inhomogeneities of the sample the local Dirac point
also changes its position. Such behavior is in a striking contrast with
the surface pinning of the Fermi level in the most of three-dimensional semiconductors.

The measurements done in Ref.~\onlinecite{He2015APL}  showed that the work
function of uniaxially strained graphene
increases by $\SI{0.16}{eV}$ under a $7\%$ strain, i.e with the
relative rate $d \work/d \varepsilon = \SI{2.17}{eV}$.
For a doped strained graphene the work function
can be expressed as $\work = \work_D^\varepsilon + \work_F^\varepsilon $, viz. a sum
of the energy intervals between the local vacuum energy level and the Dirac point, $\work_D^\varepsilon$, defined above,
and the Dirac point and Fermi level, $\work_F^\varepsilon = \cED^\varepsilon - \cEF^\varepsilon $,
as shown in Fig.~\ref{fig:1}~(b).
Accordingly, the rate of change of the work function is
\begin{equation}
\label{WF-rate}
\frac{d \work}{d \varepsilon} = - \alpha - \frac{d \EF^\varepsilon }{d \varepsilon} ,
\end{equation}
where $\EF^\varepsilon = \cEF^\varepsilon - \cED^\varepsilon$ is
the Fermi energy  counted with respect to Dirac point $\cED^\varepsilon$. At that, the quantity $\EF^\varepsilon$
is also strain dependent.

Let us estimate the position of the Fermi level, $\EF^\varepsilon$, with respect to the Dirac point
in the CVD graphene samples studied  in Ref.~\onlinecite{He2015APL}.
Using the DOS of the pristine graphene (\ref{DOS-atom-cell})
one can obtain the relationship between the number of carriers per atom  $N_c$
($N_c >0$ for the electron doped and $N_c <0$ for the hole doped)
and Fermi energy of noninteracting Dirac fermions \cite{foot}
in uniaxially strained clean graphene
\begin{equation}
\Nc = \frac{\mathrm{sgn}(\EF^\varepsilon) (\EF^\varepsilon)^2}{ W^2_\varepsilon}.
\label{eqn:n-carriers}
\end{equation}
Then, assuming that the number of carriers per atom  $N_c$ is fixed, we arrive at the following
expression for the strain induced shift of the Fermi energy with respect to the Dirac point
\begin{equation}
\EF^\varepsilon = (\cEF - \cED) \sqrt{ 1 - \beta \varepsilon} ,
\label{eqn:efermi-epsilon}
\end{equation}
where $\cEF$ is the Fermi energy of unstrained graphene.
Since the slope of the DOS (\ref{DOS-area}) increases,
the Fermi level goes towards the Dirac point to
accommodate the same number of carriers. To
illustrate this behavior we show the aforementioned shift in Fig.~\ref{fig:1}~(a).
The position of the Dirac point in Fig.~\ref{fig:1}~(a) is assumed to be independent of strain and
hole doping, $(\cEF - \cED) < 0$, is considered.

Taking the given above estimate $\alpha \approx \SI{-2.5}{eV}$,
one finds from Eq.~(\ref{WF-rate}) the value of the derivative
$d \EF^\varepsilon /d \varepsilon \approx
\SI{0.33}{eV}$. Now, using Eq.~(\ref{eqn:efermi-epsilon}) we obtain that
$\cEF - \cED = -\SI{0.22}{eV}$ (hole doping) for $\beta=3$. This value is surprisingly close to the absolute value $|\cEF - \cED| = \SI{0.23}{eV}$ extracted from the optical spectroscopy measurements
in CVD graphene \cite{Kuzmenko20172DMat}.

Thus, when the number of carriers in a graphene sample is fixed, not
only the Dirac point moves to the left, but also
the interval $|\EF^\varepsilon| =|\cEF^\varepsilon - \cED^\varepsilon|$
diminishes under the strain, as it is shown in Fig.~\ref{fig:1}~(a).
Contrary to that, the spectroscopic measurements \cite{Kuzmenko20172DMat} indicate that the
Drude weight and the corresponding quantity $ |\EF^\varepsilon|$ (see Sec.~\ref{sec:scattering-rate} below)
are essentially strain independent.
This observation  can be easily understood since for the
fixed number of carriers $1\%$ strain yields the decrease of the interval
$|\EF^\varepsilon|$ by a mere $\SI{3.5}{meV}$, which is well within the experimental
error of Ref.~\onlinecite{Kuzmenko20172DMat}.
Whereas if the position of the Fermi level $\cEF^\varepsilon$ is assumed to be fixed,
the strain-induced shift of the Dirac point would result in $\SI{25}{meV}$ decrease of
the Drude weight for $1\%$ strain and hole doping, which does not show up.

In what follows, we consider two cases:
(i) $|\EF^\varepsilon| = \mbox{const}$ that corresponds to the constant Drude weight
observed in the experiment in Ref.~\onlinecite{Kuzmenko20172DMat}
and (ii) the isolated sample with fixed number of carriers, taking into account the small
drift of the value $|\EF^\varepsilon| $ described by Eq.~(\ref{eqn:efermi-epsilon}).

\subsection{Strain effect on impurities}

In addition to the change of the host lattice parameters that takes place under a strain,
we have to take into account the analogous change of parameters describing an impurity.
In the Lifshitz model, it is natural to assume that the impurity potential $\VL$ is also strain-dependent.
Thus, one has to add a term similar to the scalar potential described in Eq.~(\ref{eqn:deformation-potential}), with a distinct parameter $\alpha'$ signifying the strain effect on impurity sites.
In our treatment of the Lifshitz model, we will neglect this discrepancy,
and use the same value of the deformation potential for both host and impurity sites as a zero-order approximation.

Nonetheless, we will examine the role of the deformation potential more carefully in our treatment
of the Fano-Anderson model (\ref{eqn:hamiltonian-fano}). To do this consistently, one has to specify
how the impurity hopping parameter $\thyb$ and the potential on the impurity $U_{\alpha l}^{\mathrm{imp}}$
change under the strain.
To our best knowledge, there are no reliable data on strain dependence of these quantities.
Thus, we cannot justifiably estimate the rate of change of the impurity parameters, $\thyb$ and $U_{\alpha l}^{\mathrm{imp}}$,
under a strain in the same way as we did for the host atoms.

Still, we intend to consider the strain dependence of the Fano-Anderson model parameters
in the following way. We will assume that the hopping parameter $\thyb$ does not change under the strain.
To describe the change of the impurity potential, we will use the linear law similar to the one used for the host potential [cf. Eq.~(\ref{eqn:deformation-potential})]:
\begin{equation}
U_{\alpha l}^\mathrm{imp} =\cED + \e_0 + \alpha^\mathrm{imp} \varepsilon.
\end{equation}
Here $\e_0$ is the difference between the host site and impurity site potential for zero strain, and $\alpha^\mathrm{imp}$ is a deformation potential  for impurities which in general differs from $\alpha$.

\section{Methods}
\label{sec:formalism}

As we mentioned in Sec.~\ref{sec:deformation-potential} below
Eq.~(\ref{WF-rate}), it is rather convenient to count the Fermi energy from
the  strain-dependent Dirac point, $\cED^\varepsilon$. Thus
we introduce the notation
\begin{equation}
\label{eqn:relative-energy}
\e = \cE - \cED^\varepsilon
\end{equation}
and use it in what follows.

\subsection{Diagonal element of the host Green's function}
\label{sec:host-GF}

The diagonal element of the host Green's function (GF)
\begin{equation}
\label{GF-host}
\hat{\mathrm g}^\varepsilon (E) = [\hat{ \mathrm I}  (E + \cED^\varepsilon) - \hat{\mathrm H}_{\mathrm{host}}]^{-1}
\end{equation}
in the site representation is necessary to proceed with the calculations.
Here, $\hat{\mathrm H}_{\mathrm{host}}$ is the Hamiltonian
(\ref{eqn:hamiltonian_host}) written in the form of $N \times  N$ matrix and $\hat{\mathrm I}$ is
the unit $N \times  N$ matrix, respectively. Since the inversion symmetry is preserved, the diagonal matrix elements
of the GF are identical on both sublattices, $\hat{\mathrm g}_{lAlA}(E) = \hat{\mathrm g}_{lBlB}(E)$.

The imaginary part of the diagonal element of the retarded GF, $g_{0}^\varepsilon (E) = \hat{\mathrm g}_{lAlA}(E + i 0)$,
is related to the DOS per atom as follows
\begin{equation}
\mathrm{Im} g_0^\varepsilon (E) = - \pi D_\varepsilon (E) .
\end{equation}
Its real part can be restored using the Kramers-Kronig relation
\cite{Skrypnyk2007FNT}.
For low energies $E$ defined by Eq.~(\ref{eqn:relative-energy})
the final expression for the diagonal part of the host Green's function acquires the following form:
\begin{equation}\label{g0}
g_{0}^\varepsilon (\e)  = \frac{\e}{W^2_\varepsilon}
\ln\left(\frac{\e^2}{W^2_\varepsilon}\right) - i \frac{\pi |\e|}{W^2_\varepsilon}, \qquad |E| \ll W_0.
\end{equation}

\subsection{Consideration of Lifshitz and Fano--Anderson models}
\label{sec:Lifshitz-formalism}

Dealing with the impurity problem, one may rely on the conventional analytic approach
developed for a substitutional binary alloy  \cite{Elliott1974RMP,Lifshits1988book}
(see also review \cite{Skrypnyk2018FNT} and Ref.~\onlinecite{Pershoguba2009PRB}, where graphene was studied).
Let us treat the case of the Lifshitz model first.

A perturbed GF for the Lifshitz model can be defined for each arrangement
of impurities as follows:
\begin{equation}
\label{GF-Lifshitz}
 \mathcal{\hat G}^\varepsilon (E) = [\hat{\mathrm I} (E + \cED^\varepsilon) - \hat{\mathrm{H}}_\mathrm{host}
- \hat{\mathrm{H}}_{\mathrm{L}} ]^{-1} .
\end{equation}
The averaged over impurity distributions GF of
the disordered system $\hat{\mathrm G}(E) =\langle \mathcal{\hat G} (E)
\rangle$ is related to the host GF  by the Dyson equation
\begin{equation}
\label{eqn:Dyson}
\hat{\mathrm G}^\varepsilon   = \hat{\mathrm g}^\varepsilon  + \hat{ \mathrm g}^\varepsilon   \hat \Sigma  \hat{\mathrm G}^\varepsilon ,
\end{equation}
where $\hat \Sigma $ is the self-energy operator with the index $\varepsilon$
suppressed for brevity.
We omit scatterings from pairs and larger groups of impurities, i.e., neglect cluster effects.
The self-energy is diagonal in this approximation, i.e.,  $\hat \Sigma = \varSigma \hat{ \mathrm I}$.
Thus, the solution of the Dyson equation can be expressed as
\begin{equation}
\hat{\mathrm G}^\varepsilon (E) =\hat{\mathrm g}^\varepsilon [E - \varSigma(E)],
\end{equation}
which corresponds to the renormalization of the host GF.

In this work we employ two approximations for the solution of  Eq.~(\ref{eqn:Dyson}).
The first one is the average $T$-matrix approximation (ATA) \cite{Ehrenreich1976}. In this approximation,
the self-energy function acquires the form
\begin{equation}
\label{eqn:sigma-ata-lifshitz}
\varSigma_{\mathrm{ATA}} (E) = \frac{c V_L}{1 - (1 - c) V_L g_0^\varepsilon(E)}
\end{equation}
with $g_0^\varepsilon(E)$ given by Eq.~(\ref{g0}).

Apart from the ATA, we employ the coherent-potential approximation (CPA) \cite{Elliott1974RMP}.
In this approximation, the self-energy is expressed as a solution of the following equation:
\begin{equation}
\label{eqn:sigma-cpa-lifshitz}
\varSigma_{\mathrm{CPA}} (E)= \frac{c V_L}{1 -
[V_L-\varSigma_{\mathrm{CPA}} (E)] g_0^\varepsilon [E - \varSigma_{\mathrm{CPA}}(E)]}.
\end{equation}

It can be shown \cite{Skrypnyk2013JPCM} that for the Fano-Anderson model, the host part of the Green's function has
the same form as in the Lifshitz model, with the impurity potential $\VL$ replaced by the energy-dependent effective potential
\begin{equation}
\label{eqn:v-effective-alpha}
V_\varepsilon(\e) = \dfrac{|\thyb|^2}{\e - E_0 + \Delta \alpha \, \varepsilon},
\end{equation}
where $\Delta \alpha = \alpha - \alpha^\mathrm{imp}$ is the difference between the deformation potential  on host and impurity atoms.
To obtain the expressions for the self-energy analogous to (\ref{eqn:sigma-ata-lifshitz}) and (\ref{eqn:sigma-cpa-lifshitz})  for the Fano-Anderson model, one should substitute the impurity potential $V_L$ by the effective potential $V_\varepsilon(E)$ defined in Eq.~(\ref{eqn:v-effective-alpha}).

\subsection{Relationship between the Drude width and the impurity scattering rate}
\label{sec:scattering-rate}

Let us now discuss the link between the self-energy
$\varSigma$
and the parameters extracted from the spectroscopy measurements
\cite{Kuzmenko20172DMat,Yu2016PRB}.
We assume that the Drude peak in ac conductivity has the Lorentzian
shape, viz.
\begin{equation}
\label{conductivity}
\mbox{Re} \sigma(\omega) = \frac{\mathcal{D} (E_F)}{\pi}
\frac{\Gamma_{\mathrm{opt}} (E_F)}{\omega^2 + \Gamma_{\mathrm{opt}}^2
(E_F)}.
\end{equation}
Here $\mathcal{D}$ is the Drude spectral weight and $\Gamma_{\mathrm{opt}} = 2
\Gamma_{\mathrm{tot}}$ is the Drude peak width (with
$\Gamma_{\mathrm{tot}}$ being
the total single particle scattering rate).
One of the advantages of
the infrared spectroscopy,
as compared to the dc transport measurements, is that it allows to
extract both the Drude spectral weight
and the Drude peak width.
The uniaxial strain makes the Drude weight anisotropic
\cite{Pellegrino2011HP} and  sensitive to the anisotropy of the
Fermi velocity (see below Eq.~(\ref{Dirac-strain})).
However, presence of a substrate does not allow to measure
the intrinsic dichroism of graphene \cite{Kuzmenko20172DMat}. Thus,
in Ref.~\onlinecite{Kuzmenko20172DMat}, the averaged over strain directions Drude weight
was considered.
In the absence of impurities
this weight is merely proportional to the  Fermi energy counted with
respect to the Dirac point,
\begin{equation}
\label{eqn:d-free}
\mathcal{D} (\EF^\varepsilon) = \frac{e^2}{\hbar^2} |\EF^\varepsilon|
\end{equation}
that allows one  to find out the absolute value of $\EF^\varepsilon$
and its strain dependence \cite{Kuzmenko20172DMat}.

In the general case, the simple relationship (\ref{eqn:d-free}) is violated
by the presence of defects.
However, when the concentration of defects is small, $|\varSigma(\EF)| \ll |\EF|$, another
rather simple equation for the Drude weight,  which
involves the quasiparticle self-energy, is valid \cite{cmp2018}
\begin{equation}
\label{eqn:d-simple}
\mathcal{D}(\EF) = \frac{e^2}{\hbar^2} [\EF -
\mbox{Re}\varSigma(\EF)], \quad |\varSigma(\EF)| \ll |\EF|.
\end{equation}
As discussed in Sec.~\ref{sec:results}, in the present work we restrict ourselves to a small
impurity concentration, $c \leq 4 \times 10^{-4}$. In this case, $| \varSigma(\EF)| \lesssim \SI{10}{meV} $,
which is indeed much less than the value $|\EF| \sim \SI{0.23}{eV}$ reported for CVD graphene \cite{Kuzmenko20172DMat}.
Furthermore, the variation of $\mbox{Re}\varSigma(\EF)$ under the strain is of the order of a few millielectronvolts,
so that in the case when $\EF^\varepsilon$ is fixed, one can also assume that the Drude weight also
remains practically constant. This situations breaks down for high-impurity concentrations,
when Eq.~(\ref{eqn:d-simple}) is not applicable (see  Ref.~\onlinecite{cmp2018})
and fixing the value of $\EF$ does not imply that the Drude weight remains constant.

According to Matthiessen’s rule, the total single-particle
scattering rate is
$\Gamma_{\mathrm{tot}} = \sum_i \Gamma_i$, where $\Gamma_i$ are the
contributions from different sources of scattering, e.g.,
short-range point defects, long-range charged impurities, acoustic
phonons, surface phonons in
the substrate and grain boundaries. Each of these contributions may
be strain dependent.
In thші work we restrict ourselves to
the effect of resonant impurity scattering on
the Drude-peak width
\begin{equation}
\label{scattering-rate}
\Gamma_{\text{opt}, \varepsilon}(\EF ) = 2 \Gamma_\varepsilon(\EF) = -
2\mbox{Im} \, \varSigma (\EF)
\end{equation}
as a function of strain $\varepsilon$ and Fermi energy, $\EF$.
The effect of small strain on the scattering rate and the Drude-peak width
can be characterized by the relative gain
\begin{equation}
\varkappa  = \frac{\partial \ln \Gamma_\varepsilon }{\partial
\varepsilon} \Big |_{\varepsilon = 0}.
\label{eqn:relative-gain}
\end{equation}

\section{Results}
\label{sec:results}

\subsection{Lifshitz model}
\label{subsec:results-lifshitz}

In this section, we present results obtained for the Lifshitz model.
It is instructive to begin our consideration with discussion of the
correspondence between
our approach and an estimate of the scattering rate from point
defects done in Ref.~\onlinecite{Kuzmenko20172DMat}.

\subsubsection{Weak scattering regime}
\label{sec:Born}

The Born weak-scattering approximation  follows from the ATA equation
(\ref{eqn:sigma-ata-lifshitz}) for small $V_L$. For $\Gamma_{\mathrm{Born}} = - \mbox{Im} \, \varSigma (E_F) $ it gives
\begin{equation}
\Gamma_{\mathrm{Born}} = - c V_L^2 \Im g_0^\varepsilon(E_F) = \pi c V_L^2 D_\varepsilon (E_F).
\label{eqn:gamma-born}
\end{equation}
Since we assumed that the impurity potential $V_L$ is independent of strain, one obtains that $\varkappa_{\mathrm{Born}} = \beta$.
This result is substantially larger than the estimate done in Ref.~\onlinecite{Kuzmenko20172DMat}, that gave
$\varkappa_{\mathrm{Born}}^\prime \approx 0.3$.
This disagreement can be understood from the fact that the potential
used in the
continuum descriptions of point defects
$V (\mathbf{r})= V_{\mathrm{pd},\varepsilon} \delta (\mathbf{r})$
contains the constant $V_{\mathrm{pd},\varepsilon}$ that was assumed
in Ref.~\onlinecite{Kuzmenko20172DMat} to be independent of strain.
This assumption would indeed result in $\varkappa_{\mathrm{Born}}^\prime
= \beta -2$.
However, since $V_L$ is assumed to be independent of strain,
the energy density $V_{\mathrm{pd},\varepsilon}$ depends on the
strain as
$V_{\mathrm{pd},\varepsilon} \simeq V_{\mathrm{pd},0} (1 +
\varepsilon)$, with  $V_{\mathrm{pd},0}$ being the
corresponding energy density without strain. This strain dependence of $V_{\mathrm{pd},\varepsilon}$ allows to recover
our estimate $\varkappa_{\mathrm{Born}} = \beta$.
As it can be seen from the above, in the weak scattering limit the self-energy is a slowly-varying linear function of energy.
\subsubsection{Large impurity perturbations}
\label{sec:lifshitz-resonance}

As was shown in Ref.~\onlinecite{Skrypnyk2006PRB}, in the Lifshitz model
a resonance  state appears for large impurity potentials.
It is known that in vicinity of the resonance energy scattering from  impurities significantly increases.
Thus, we examine below the case of strong impurity potentials.

Since we consider the hole-doped regime, $\EF < 0$, which was studied in the experiments \cite{Kuzmenko20172DMat,He2015APL},
we are mostly interested in those impurities that induce resonance states with energies located below the Dirac point.
We note that the case of electron-doped sample, $\EF > 0$, can be treated in the same fashion, but
a resonance with the positive energy is necessary to enhance the corresponding scattering rate.

The presence of the impurity resonances was numerically confirmed by \emph{ab initio} calculations \cite{Robinson2008PRL,Wehling2010PRL,Ihnatsenka2011} and observed in experiments \cite{Bostwick2009}.
The examples of such impurities are the adsorbed atoms \ch{H} and \ch{F}, hydroxyl groups \ch{OH^-}, etc.
The energy of the resonance depends on the sort of the impurity.
For example, according to  Ref.~\onlinecite{Ihnatsenka2011}
the resonances formed by \ch{H}, \ch{F}, and \ch{OH^-} are predicted to have energies $-0.07$, $-0.38$, and $\SI{-0.25}{eV}$, respectively.
However, there is a significant inconsistency between values of the resonance energy obtained by different methods.
For example, the \ch{H} resonance energy is considered to be in $\pm \SI{0.03}{eV}$ interval in Ref.~\onlinecite{Wehling2010PRL}, at $\SI{-0.07}{eV}$ in Ref.~\onlinecite{Ihnatsenka2011} and at $\SI{0.20}{eV}$ in Ref.~\onlinecite{Bostwick2009}.
In our analysis, we will not restrict ourselves to a particular sort of impurity.
Instead, we will pick a resonance energy to demonstrate the proposed mechanisms of the scattering rate enhancement
most vividly, albeit using the mentioned results as a guide.

\begin{figure}[h]
\centering
\includegraphics[width=\linewidth]{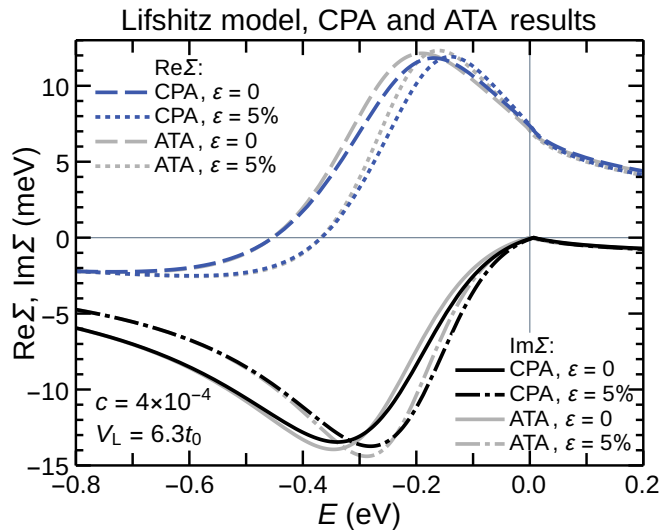}
\caption{The real and imaginary parts of the self-energy function $\varSigma(E)$ in vicinity of the Dirac point in the Lifshitz model.
The model parameters are $V_L= 6.3 t_0$, $c = 4 \times 10^{-4}$.
The convex and concave curves correspond to $\Re \varSigma(E)$ and $\Im \varSigma(E)$, respectively.
Blue lines and black lines were computed in the CPA approximation, while gray lines were
obtained  in the ATA approximation.
The dashed and the solid lines correspond to the results for zero strain, while
the dot-dashed and dotted lines are for $\varepsilon=5\%$.}
\label{fig:lifshitz-sigma}
\end{figure}
In Fig.~\ref{fig:lifshitz-sigma} we present results for the real and imaginary parts
of the self-energy function, $\varSigma(E)$, for $0$ and $5\%$ strain.
We remind that to compare results for different values of strain, we present them as functions of
the relative energy $E$ defined by Eq.~(\ref{eqn:relative-energy}), so that
for arbitrary strain the Dirac point energy is located at the origin, $E = 0$.
The impurity perturbation  $\VL = 6.3 t_0$ is chosen to give a minimum in $\Im \varSigma(E)$ approximately at
$-\SI{0.35}{eV}$. Such impurity perturbation can be attributed to the fluorine impurities.
The impurity concentration is chosen to be $c = 4 \times 10^{-4}$.
The self-energy function was calculated both in the ATA [Eq.~(\ref{eqn:sigma-ata-lifshitz})]
and the CPA [Eq.~(\ref{eqn:sigma-cpa-lifshitz})] approximations, respectively.
One can see that both approximations give rather similar results for the given choice of the parameters.
We observe a peak in the real part and a dip in the imaginary part of the self-energy below the Dirac point, $E = 0$.
A more careful examination reveals that the real part has an inflection point at the same energy in which the minimum
of $\Im \varSigma(E)$ is located.
As the strain is applied, the inflection point and the dip get closer to the Dirac point.

To identify the described above features in the self-energy function, we now investigate the ATA approximation (\ref{eqn:sigma-ata-lifshitz}) analytically.
Neglecting the concentration $c$ near the unity in the denominator,  we obtain the imaginary part
of the self-energy function in the following form:
\begin{equation}
\Im \varSigma(E) \approx
\dfrac{-c \VL^2 \pi |E| / W^2_\varepsilon}
{\left[1- \dfrac{\VL E}{W^2_\varepsilon} \ln
	\left(\dfrac{E^2}{W^2_\varepsilon}\right)  \right]^2 +
\left[\dfrac{\pi \VL E}{W^2_\varepsilon} \right]^2 } .
\label{eqn:sigma-ata-lifshitz-simplified}
\end{equation}
This fraction rapidly decreases as the first term in the denominator vanishes at some energy $\Er$,
which is yielded by the Lifshitz equation:
\begin{equation}
\label{eqn:lifshitz-equation}
1- \dfrac{\VL \Er}{W^2_\varepsilon} \ln
	\left(\dfrac{\Er^2}{W^2_\varepsilon}\right)
	= 0.
\end{equation}
As the perturbation $|\VL|$ increases, the value $\Er$ converges to the position,
where the function  $\Im \varSigma(E)$ has the minimum that corresponds to the
maximum in the full DOS, which includes the effect of impurities.
Thus, the energy $\Er$ marks the position of the impurity resonance in the Lifshitz model.
We will refer to $\Er$ as the resonance energy, even though for moderate values of $\VL$ it slightly
deviates from the minimum in the self-energy.

In the Lifshitz model, the location of the resonance relative to the Dirac point is opposite to the sign
of the impurity perturbation: $\mathrm{sgn}(\Er) =- \mathrm{sgn}(\VL)$.
This property is inherent in general spectra consisting of two symmetric bands touching each other
in a finite number of points \cite{Skrypnyk2007FNT}.

The resonance energy shifts with varying the impurity perturbation $\VL$.
To determine the direction of this shift, we took a partial derivative over $\VL$ of the  Lifshitz equation (\ref{eqn:lifshitz-equation}):
\begin{equation}
\label{eqn:lifshitz-derivative}
\dfrac{\partial \Er}{\partial \VL} = \frac{1}{2} \left(\dfrac{ W_\varepsilon}{\VL}\right)^2  \left[\ln \frac{ W_\varepsilon}{|\Er|} - 1\right]^{-1}.
\end{equation}
This expression is positive for $|\Er| <  W_\varepsilon/\mathrm{e}$, where $\mathrm{e}$ is the base of the natural logarithm.
As long as we consider strong potentials, one can see that as $|\VL|$ increases,
the resonance energy  shifts towards the Dirac point energy in either $\Er >0$ or $\Er < 0$ case.

A similar shift of the resonance energy occurs when uniaxial strain is applied.
As we stretch the sample, the bandwidth $ W_\varepsilon$ decreases [see Eq.~(\ref{eqn:w-epsilon})].
The bandwidth acts as a scaling parameter in  the solution of the Lifshitz equation (\ref{eqn:lifshitz-equation}).
Indeed, this solution  can be written in the  form  $\Er /  W_\varepsilon =  f(\VL /  W_\varepsilon)$,
where $f(x)$ is a function of the dimensionless potential $\VL/ W_\varepsilon$.
The latter determines the position of the resonance in units of $W_\varepsilon$.
As discussed below Eq.~(\ref{eqn:lifshitz-derivative}), the resonance is getting closer to
the Dirac point for stronger impurity potentials.
The decrease of the bandwidth leads to the two consequences, viz.,
trivial rescaling of $\Er$ and an increase of $\VL/ W_\varepsilon$.
Both of them result in the decrease of the absolute value of the resonance energy $\Er$.

To verify this result, we have also obtained an exact solution of the
Lifshitz equation (\ref{eqn:lifshitz-equation}).
The solution can be written in terms of the Lambert function which is
defined as a root $F(z)$ of the equation $z= F \exp(F)$ \cite{Corless1996ACM}.
The Lambert  function has two branches $F_0(z)$ and
$F_{-1}(z)$, which represent two single-valued solutions of the equation.
Specifically, we use $F_{-1}$ branch that satisfies the condition $|F_{-1}(x)| > 1$
for $x < 0$ and corresponds to the resonance inside the band, $|\Er| < W_\varepsilon$.
Then the solution of Eq.~(\ref{eqn:lifshitz-equation}) can be written in the form
\begin{equation}
\Er = \frac{ W_\varepsilon^2}{2 \VL} \frac{1}{F_{-1}
\left( -\tfrac{W_\varepsilon}{2 |\VL|} \right)}.
%\Er = \frac{ W_\varepsilon^2}{2 \VL} \frac{1}{F_{-1} ( W_\varepsilon/2\VL)}.
\label{eqn:lifshitz-eqn-solution-exact}
\end{equation}
In accordance with the above-mentioned arguments, the resonance energy shifts toward the Dirac point,
$E = 0$, as the strain $\varepsilon$ increases.

Calculations performed for different values of $\VL$ give results resembling those shown in Fig.~\ref{fig:lifshitz-sigma}.
For a reference, the values of $\VL$ in the range between $15$ and $\SI{30}{eV}$ produce
the resonance energies in the range between $-0.40$ and $\SI{-0.20}{eV}$.
In addition to the resonance position moving closer to the Dirac point for larger
$|\VL|$, the resonance peak also gets narrower.
As for the impurity concentration $c$, it mainly acts as a linear scale for absolute value
of the self-energy function. Increasing $c$ does not affect the positions of the extrema in
the ATA approximation, and in the CPA approximation the results indicate a slight shift
for high concentrations.
In the limit of small impurity concentration, $c \to 0$, both approximations give the same results.

\subsection{Fano-Anderson model}
\label{subsec:results-fano}

Although the Lifshitz model does reasonably well describe substitutional impurities
and is capable of generating a resonance state near the Dirac point for large values
of the impurity perturbation $\VL$, we are going to proceed to the Fano-Anderson model analysis for several reasons.

First, the Lifshitz model has only one adjustable parameter, $\VL$.
The real impurities are expected to alter more than one matrix element of the host Hamiltonian.
Such can be the hopping integrals and the potential energies on the lattice sites in the neighborhood of the impurity.
Generally, we expect a model with extra parameters to be able to provide for more accurate results.
In this regard, the Fano-Anderson model is preferable, as it has two parameters, $E_0$ and $\thyb$.

Second, we need a model to describe adsorbate impurities.
This is the type of point defects we expect to be present in noticeable quantities in graphene
samples grown by conventional methods on substrates.
It is widely known that the Fano-Anderson model is capable of describing, at least qualitatively,
adsorbates deposited on a graphene sheet: atoms, molecules, free radicals, etc.

Third, to obtain the impurity resonance in the Lifshitz model, we had to assign unrealistically large values of
the impurity perturbation $\VL$. In contrast, the Fano-Anderson model is free of this deficiency.
It is capable to provide resonances with energy and width nearly identical to that of the Lifshitz model,
but the respective  values of the parameters $\thyb$ and $E_0$ are of the same order as $t_0$.

Hereby we will describe how Fano-Anderson impurities form a Drude peak width that is highly sensitive to applied strain.
If we want to provide a description for a specific sort of impurities, we have to assign realistic values to
the model parameters, $\thyb$ and $E_0$.
Such values can be obtained by fitting to \emph{ab initio} results, as was done in Refs.~\onlinecite{Robinson2008PRL,Wehling2010PRL,Ihnatsenka2011}.
The results vary due to obvious imprecision of the fitting procedure.
After all, this model does not describe adsorbate impurities in every respect.
In our examples, we will adjust the model parameters to match a pre-defined resonance energy,
while keeping the \emph{ab initio} fits in  mind.

As we have noted previously, little is known on the response of the impurity potential to strain.
We overcome this difficulty by sweeping the impurity potential  $\alpha^\mathrm{imp}$ between two limiting values:  $\alpha^\mathrm{imp}=\alpha$ and $\alpha^\mathrm{imp}=0$.

In the first limiting case, the potential energy on impurity sites changes in concert with host sites,
so that $\alpha^\mathrm{imp} = \alpha = -\SI{2.5}{eV}$. Thus, the relative deformation potential  $\Delta \alpha = 0$,
and it is not present in the effective impurity potential  (\ref{eqn:v-effective-alpha}).

In the second limiting case, the potential on impurity sites does not change at all with the strain, $\alpha^\mathrm{imp} = 0$.
Therefore, $\Delta \alpha = \alpha = -\SI{2.5}{eV}$ is present in Eq.~(\ref{eqn:v-effective-alpha}).
\begin{figure}[h]
\centering
\includegraphics[width=1\linewidth]{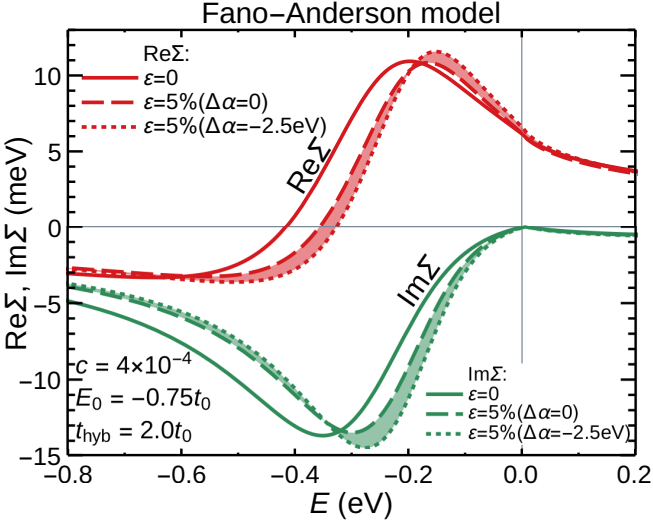}
\caption{
The real (upper red curves) and imaginary (lower green lines)
parts of the self-energy $\varSigma(E)$ for the Fano-Anderson model calculated in the CPA approximation
with $E_0 = - 0.75 t_0$, $\thyb = 2 t_0$, $c = 4 \times 10^{-4}$ for two values of strain $\varepsilon = 0$ and $\varepsilon = 5\%$.
The solid lines show results for zero strain, the dashed lines show results for $\varepsilon = 5 \%$ with the same deformation potential for host and impurity atoms ($\Delta \alpha = 0$), and the dotted lines show results for $\varepsilon = 5 \%$ with no impurity deformation potential ($\alpha^{\mathrm{imp}} = 0$, or $\Delta \alpha = -\SI{2.5}{eV}$).
Results obtained with $\alpha^{\mathrm{imp}}$  varying between these two limiting values span the shaded areas.}
\label{fig:fano-cpa-re-im-sigma}
\end{figure}
In Fig.~\ref{fig:fano-cpa-re-im-sigma}, we show real and imaginary parts of the self-energy function, $\Re \varSigma(E)$, and $\Im \varSigma(E)$, calculated in the CPA approximation for $0$ and $5\%$ strain.
The parameter $\thyb = 2 t_0$ is chosen to conform with the general estimate provided in Ref.~\onlinecite{Wehling2010PRL}.
While this choice is not the only possible one, as the values for specific impurities may differ significantly (like in Ref.~\onlinecite{Ihnatsenka2011}), we consider it as a reasonable compromise that provides a description of strongly bound impurities.
The other parameter, $E_0 = - 0.75 t_0$, is tuned to match with the resonance energy   $\Er =-\SI{0.35}{eV}$,
in analogy to Fig.~\ref{fig:lifshitz-sigma}. This is surprisingly  close to the parameters for  fluorine
obtained in the recent preprint \cite{Wellnhofer2019}.
Whereas these results resemble those presented for the Lifshitz model in Fig.~\ref{fig:lifshitz-sigma},
the values of  $\thyb$ and $E_0$ are  not as excessively large as $\VL = 6.3 t_0$ used there.

To demonstrate a possible contrast between the results with and without the impurity deformation potential, in Fig.~\ref{fig:fano-cpa-re-im-sigma} the areas between the two lines with $\Delta \alpha = 0$ (dashed lines)
and $\Delta \alpha = -\SI{2.5}{eV}$ (dotted lines) are shaded.
The apparent difference suggests that the effect of the impurity deformation potential
could be as significant as the one resulted from variation of the bandwidth.

The obtained results are quite similar to the ones we have seen for the Lifshitz model.
Let us clarify how far this comparability extends.
\begin{figure}[h]
\centering
\includegraphics[width=1\linewidth]{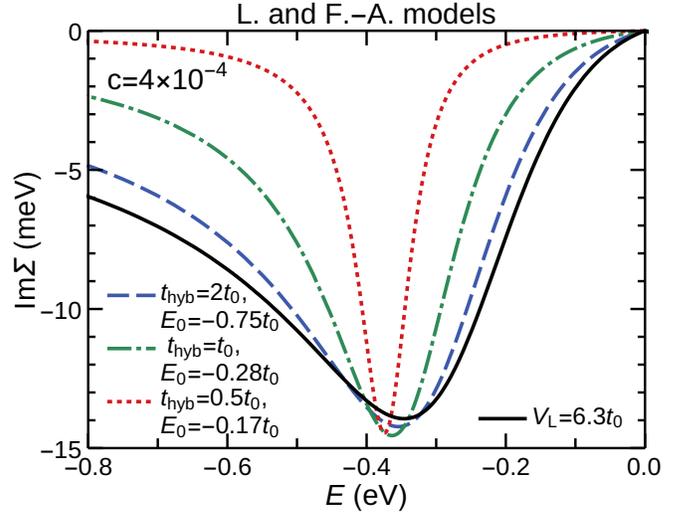}
\caption{Imaginary part of the self-energy $\Im \varSigma(E)$ calculated for the Lifshitz model (black solid line)
and for the Fano-Anderson model (other curves).
The impurity concentration $c = 4 \times 10^{-4}$ is the same for all curves.
The parameter $\VL = 6.3 t_0$  for the Lifshitz model. The parameters for the Fano-Anderson model are:
$\thyb = 2 t_0$, $E_0 = -0.75 t_0$ (blue dashed line); $\thyb =  t_0$, $E_0 = -0.28 t_0$ (green dotted-dashed line);
$\thyb = 0.5 t_0$, $E_0 = -0.17 t_0$ (red dotted line). The zero strain is considered.
}
\label{fig:lifshitz-fano-im-sigma}
\end{figure}
In Fig.~\ref{fig:lifshitz-fano-im-sigma}, we present $\Im \varSigma$ calculated both for
the Lifshitz model (black solid line) and for the Fano-Anderson model (other curves) at zero strain.
To make a proper comparison, the position of the resonance for the Lifshitz model  with $\VL = 6.3 t_0$
has to be matched with the positions of the minima of the self-energy function obtained for the Fano-Anderson model.
We fulfill this goal by using the following procedure.
To cover a wider range of the model parameters we select the impurity hopping parameter $\thyb$ as multiples of $0.5 t_0$.
Then we adjust the difference between the host site and impurity
site potential for zero strain, $E_0$, to achieve the chosen value of the resonance energy. The impurity concentration
$c = 4 \times 10^{-4}$ is fixed for all curves.

We observe in Fig.~\ref{fig:lifshitz-fano-im-sigma} that for weakly bound impurities with $\thyb = 0.5 t_0$
(red dotted curve) the dip in  $\Im \varSigma(E)$ becomes significantly narrower.
For strongly bound impurities with $\thyb = 2 t_0$ (blue dashed line), the Fano-Anderson model
 produces the results similar to the Lifshitz model as expected, because
both $|\thyb|$ and $|E_0|$ are significantly larger than the energy of the resonance.
In this case the function $V(E)$ given by Eq.~(\ref{eqn:v-effective-alpha}) has a weak dependence on
$E$ when the energy is located in the vicinity of the Dirac point.

In what follows, we restrict ourselves to the case $\thyb = 2 t_0$ in which
the Fano-Anderson model is expected to yield results comparable to the ones obtained in the
Lifshitz model, so that the strain dependence of the the self-energy function has similar character.
Specifically, the resonance energy $\Er$ is determined by Eq.~(\ref{eqn:lifshitz-equation}),
with $\VL$ substituted by $V_\varepsilon(E)$:
\begin{equation}
\label{eqn:lifshitz-equation-fano-model}
\Er
	= \frac{E_0 - \Delta \alpha \varepsilon}{1 - \dfrac{|\thyb|^2 }{W^2_\varepsilon} \ln
		\left(\dfrac{\Er^2}{W^2_\varepsilon}\right)} .
\end{equation}
In the case of strongly bound impurities, we can neglect the first term in the denominator, and this equation acquires the same form as Eq.~(\ref{eqn:lifshitz-equation}).
Thus, we can expect the resonance energy to behave under strain in the same way as in the Lifshitz model. In addition to the strain-induced shrinking of the bandwidth (\ref{eqn:w-epsilon}), the dimensionless effective potential $ -|\thyb|^2/ (E_0 - \Delta \alpha \varepsilon) W_\varepsilon$ can be further amplified by the relative deformation potential $\Delta \alpha$. Both mechanisms result in a shift of the resonance energy toward the Dirac point.

The contribution of resonant impurity scattering to the Drude-peak width $\Gamma_{\text{opt}, \varepsilon}(\EF)$
is proportional to $\Im \varSigma$ taken at Fermi energy, $\EF$ [see Eq.~(\ref{scattering-rate})].
The sharp profile of the self-energy function makes $\Gamma_\varepsilon$ very sensitive to the position of the Fermi level.
The latter is, in fact, determined by all possible sources of excess charges. Those can be induced by electrostatic
doping and all sorts of defects. While resonance impurities contribute to the charge imbalance, they do not necessarily
determine it.

For instance, let us examine whether resonance impurities alone can be responsible for the value $\EF = \SI{-0.23}{eV}$ reported in Ref.~\onlinecite{Kuzmenko20172DMat}. For simplicity, we will assume that each impurity removes one electron from the valence band.
In calculating Figs.~\ref{fig:lifshitz-sigma}, \ref{fig:fano-cpa-re-im-sigma}, \ref{fig:lifshitz-fano-im-sigma},
we had adjusted the impurity concentration $c$ to get the value $\Gamma_{\text{opt}, \varepsilon}(\EF) \approx \SI{15}{meV}$
as reported  in \cite{Kuzmenko20172DMat}.
The resulting value, $c = 4 \times 10^{-4}$, should be compared to the carrier imbalance per atom, $\Nc$.
If $c > |\Nc|$, then it means that the number of impurities is greater than the number charge donors.

To estimate the carrier imbalance per atom $\Nc$, one can use Eq.~(\ref{eqn:n-carriers}).
For $\EF = \SI{-0.23}{eV}$ and $\varepsilon = 0$, we get the negative imbalance, i.e.,
holes, with $|\Nc| = 1.2 \times 10^{-3}$. This concentration is three times larger than $c = 4 \times 10^{-4}$.
It allows us to state that the charge doping in this particular case was caused by some
other mechanism.

\begin{figure}[h]
\centering
\includegraphics[width=1\linewidth]{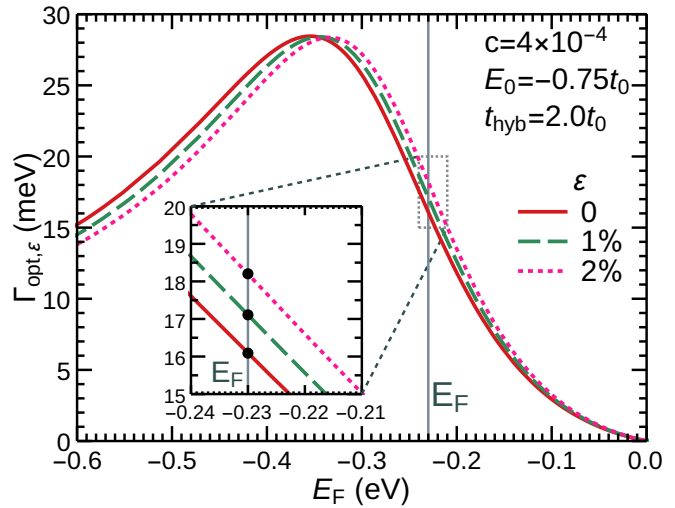}
\caption{The Drude peak width, $\Gamma_{\text{opt}, \varepsilon}
= -2\Im \varSigma(\EF)$, as a function of the Fermi energy $\EF$,
calculated for the strain $\varepsilon =0,1,2\%$ in the Fano-Anderson model.
Inset: Zoom-in  the vicinity of $\EF = \SI{-0.23}{eV}$ which shown as the vertical line.
The model parameters are the following: $c = 4\times 10^{-4}$, $E_0 = - 0.75 t_0$, $\thyb = 2 t_0$,
$\Delta \alpha = 0$.
}
\label{fig:gamma-resonance-illustrative}
\end{figure}

As discussed  in Sec.~\ref{sec:scattering-rate} (see also  the end of Sec.~\ref{sec:deformation-potential}),
the case $|\EF^\varepsilon| = \mbox{const}$ approximately corresponds to the constant Drude weight
when impurity concentration is small enough, $c \leq 4 \times 10^{-4}$.
In Fig.~\ref{fig:gamma-resonance-illustrative} we present $\Gamma_{\text{opt}, \varepsilon}$ calculated in the ATA
approximation as the function $\EF^\varepsilon$ for three increasing values of the strain $\varepsilon=0,1,2 \%$.
The parameters $\thyb = 2 t_0$ and $E_0 = -0.75 t_0$ coincide with those of Fig.~\ref{fig:fano-cpa-re-im-sigma}.
The same deformation potential for host and impurity atoms is chosen, $\Delta \alpha = 0$,
so that the shift of the resonance energy is solely caused
by the decrease of the bandwidth.
One can see in Fig.~\ref{fig:gamma-resonance-illustrative} how this shift amplifies the Drude width.

While the relative change of the resonance energy due to the strain is small compared to its absolute value
in the absence of strain, the increase in $\Gamma_{\text{opt}, \varepsilon}$ can be quite substantial.
This is a consequence of the acute steepness of $-2\Im \varSigma(E)$ function
that is reached on the half-width of the peak.
The large slope is intrinsic to the impurity resonance.
This feature is absent in the weak-scattering case, in which the resonance cannot be seen, and
the resulting expression for the Drude width dependence on the Fermi energy
is a slowly varying function with a nearly uniform moderate slope [see Eq.~(\ref{eqn:gamma-born})].

The extracted from the experimental data \cite{Kuzmenko20172DMat} value $\EF = \SI{-0.23}{eV}$ is shown as the
vertical line. The points at intersection of this line by the  curves $\Gamma_{\text{opt}, \varepsilon}$ give
values  of the Drude width that correspond  to different values of strain $\varepsilon$.
The inset in Fig.~\ref{fig:gamma-resonance-illustrative} zooms in the area with
the three mentioned intersections  for $\varepsilon=0,1,2 \%$.
One can see that the dependence of $\Gamma_{\text{opt}, \varepsilon}$ on strain is linear, because
the intersection points are equally spaced for  equal increments of the relative strain $\varepsilon$.
The linear regime holds when the values of $\EF$ fell on the half-width of the resonance peak.
It persists for values of the relative strain up to $\sim 10 \%$.

\begin{figure}[h]
\centering
\includegraphics[width=1\linewidth]{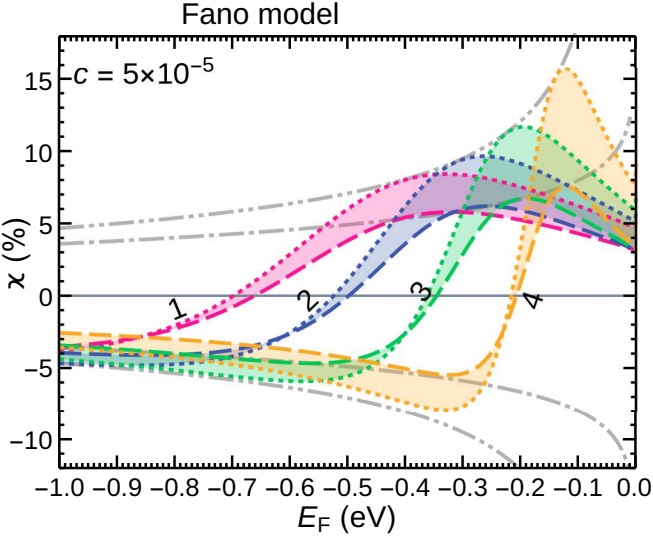}
\caption{The relative gain $\varkappa$ of the Drude width  per $1 \%$ of
strain as a function of $\EF$ for the Fano-Anderson model. The calculation is done assuming
a constant strain independent $\EF$.
All curves are for $\thyb = 2 t_0$ and $c = 5 \times 10^{-5}$.
Four sets of the curves differ by the value of $E_0$:
1, $E_0 = -1.25 t_0$ (purple lines);
2, $E_0 = - t_0$ (blue lines);
3, $E_0 = -0.75 t_0$ (green lines);
4, $E_0 = -0.5 t_0$ (orange lines).
The dashed lines show the results with the same deformation potential for host and impurity atoms ($\Delta \alpha = 0$)
and and the dotted lines are for no impurity deformation potential ($\alpha^{\mathrm{imp}} = 0$, or $\Delta \alpha = -\SI{2.5}{eV}$).
The areas between each pair of curves are shaded in same  color tone as
the lines. The dotted-dashed and double-dotted-dashed gray lines are obtained on the base of analytical estimates
for  the upper and lower bounds of the gain function for the cases
$\alpha^\mathrm{imp} = \alpha$ and $\alpha^\mathrm{imp} =0$, respectively
(see the main text for the explanation).}
\label{fig:fano-kappa-efermi}
\end{figure}

Since the Drude width scales linearly with the impurity concentration $c$,
it is convenient to consider its relative gain, which per $1\%$ of the strain reads as
\begin{equation}
\varkappa = \left. \frac{\Gamma_\varepsilon - \Gamma_0}{\Gamma_0 \varepsilon} \right|_{\epsilon=1\%}.
\label{eqn:relative-gain-1}
\end{equation}
Figure~\ref{fig:fano-kappa-efermi} shows $\varkappa$ as a function of $\EF$ calculated for the Fano-Anderson
model in the CPA approximation.
Here it is assumed that the Fermi energy itself is independent of strain, $|\EF^\varepsilon| = |\EF| = \mbox{const}$.
As was already discussed, this case approximately corresponds to the strain independent Drude weight.
The results are shown for the four sets of the model parameters, with $c = 5 \times 10^{-5}$ and $\thyb =
2 t_0$ for each set and $E_0 = -1.25 t_0$ (first set, purple lines); $E_0 = -t_0$ (second set, blue lines);
$E_0 = -0.75 t_0$ (third set, green lines); $E_0 = -0.5 t_0$ (forth set, orange lines).
Note that the $3$rd set (green lines) corresponds to the values of
$\thyb$ and $E_0$ already used in Figs.~\ref{fig:fano-cpa-re-im-sigma} and \ref{fig:gamma-resonance-illustrative}.
For each set of parameters, we calculated the gain $\varkappa$
with the same deformation potential for host and impurity atoms ($\alpha^\mathrm{imp} = \alpha$ or $\ $ $\Delta \alpha = 0$)
shown by the dotted-dashed lines and with $\alpha^\mathrm{imp} =0$ ($\Delta \alpha = -\SI{2.5}{eV}$)
plotted by the double-dotted-dashed lines.
The distances between these pairs of lines signify the contribution of the relative deformation potential to the effect.
The results for intermediate values of the relative deformation potential  ($-\SI{2.5}{eV}< \Delta \alpha < 0$) fall
in the shaded areas between the lines.

The four sets  shown in Fig.~\ref{fig:fano-kappa-efermi} differ from each other by the values of the
resonance energy $\Er$. It gets closer to the Dirac point as $|E_0|$ diminishes and passes from the first to the fourth set.
Respectively, the relative gain function $\varkappa(\EF)$ becomes sharper, as the resonance widths
are getting narrower.
This is true at least for $|\thyb| \gtrsim t_0$ case, in which the width of the resonance is proportional to $|\Er|$, like
in the Lifshitz model.

To study the relative gain $\varkappa (\EF)$ we have chosen the same impurity concentration
$c = 5 \times 10^{-5}$ for all curves. It is lower than the value $4 \times 10^{-4}$ that was used for computing Figs.~\ref{fig:fano-cpa-re-im-sigma}, \ref{fig:lifshitz-fano-im-sigma}, \ref{fig:gamma-resonance-illustrative}.
This choice ensures that for all values of $E_0$
the results presented in Fig.~\ref{fig:fano-kappa-efermi} will not change  for any lower impurity concentration.

With a significant increase in the concentration of impurities,  the ATA and CPA approximations start to deviate from each other.
More importantly, these approximations fail to give a valid description of the electronic spectrum at all energies
under consideration because at higher concentrations of impurities the spectrum undergoes a qualitative transformation.
It is estimated \cite{Skrypnyk2007FNT} that for a given energy of the resonance $\Er$, the spectrum transformation occurs
at the concentration  $c_{\mathrm{ST}} = 2 (\Er/W_0)^2 \ln( W_0/|\Er|)$. One can expect that the presented results
are applicable for the impurity concentration $c$ substantially lower than $c_{\mathrm{ST}}$. We note that
for the first and second sets the presented results are still valid even for higher concentration $c = 4 \times 10^{-4}$.
However,  the imposed restriction on the impurity concentration is fully justified for the fourth set
for which the last concentration is too high.

As can be seen from Fig.~\ref{fig:fano-kappa-efermi}, the presented curves for $\varkappa (\EF)$ functions have
a number of common features. Moving from the Dirac energy towards increasing $|\EF|$, we observe a sharp maximum followed
by a steep descent into negative values of $\varkappa$.
The zero of $\varkappa (\EF)$ function roughly corresponds to the resonance energy.
After crossing the resonance energy, we reach  the maximum decrease of the Drude weight.
Both the maximum and the minimum lie roughly on the half-width of the peak in $\Gamma(\EF)$.
Subsequently, we find a relatively small negative gain that vanishes slowly as $|\EF|$ moves away from the resonance energy and
further increases.

Let us stress that in the case $\alpha^\mathrm{imp} =0$ ($\Delta \alpha = -\SI{2.5}{eV}$)  the relative gain can be almost
twice as large compared with the case $\alpha^\mathrm{imp} = \alpha$ ($\Delta \alpha = 0$).
Thus, we cannot neglect this mechanism, and credit the hopping integral variation as the exclusive source of the partial Drude
width increase. Moreover, we cannot exclude the possibility that the relative deformation potential $\Delta \alpha$ is even
bigger in absolute value, in which case it gives even larger contribution to the effect.

We have made an analytical estimate for the  upper and lower bounds of the gain function.
A function $\varkappa_{\mathrm m} (\EF)$ describes the envelope function that goes through
extrema of $\varkappa (\EF)$ functions plotted for different values of $E_0$.
We distinguish four functions that connect the maxima or the minima for
the cases $\alpha^\mathrm{imp} = \alpha$ and $\alpha^\mathrm{imp} =0$.
The corresponding $\varkappa_{\mathrm m}(\EF)$ functions are shown in Fig.~\ref{fig:fano-kappa-efermi} by  the dotted-dashed
and double-dotted-dashed gray lines.

Since in the general case not only the bandwidth $W_\varepsilon$ is strain dependent, but also
the energy-dependent effective potential $V_\varepsilon (E)$, the relative gain $\varkappa (\EF)$
consists of the two terms,
\begin{equation}
\label{gain-const-Ef}
\varkappa (\EF)= \varkappa^{(W)} (\EF) + \varkappa^{(\alpha)} (\EF).
\end{equation}
Here the first term originates from the strain dependence of $W_\varepsilon$
\begin{equation}
\varkappa^{(W)} (\EF)= \left. \frac{1}{\Gamma_0} \frac{\partial \Gamma}{ \partial W_\varepsilon}
\frac{\partial W_\varepsilon}{\partial \varepsilon} \right|_{\varepsilon = 0},
\end{equation}
and the second is caused by the strain dependence of the potential $V_\varepsilon (E)$
\begin{equation}
\label{varkappa-alpha}
\varkappa^{(\alpha)} (\EF)  = \left. \frac{1}{\Gamma_0}
\frac{\partial \Gamma}{\partial V_\varepsilon} \frac{  \partial V_\varepsilon}{\partial \varepsilon} \right|_{\varepsilon = 0} .
\end{equation}
It is obvious that in the $\Delta \alpha = 0$ case, only the former term $\varkappa^{(W)} (\EF)$ contributes to the gain.

To find the maximal value for this term, it is sufficient to consider the imaginary part of the
self-energy for the Lifshitz model in the ATA approximation,  Eq.~(\ref{eqn:sigma-ata-lifshitz-simplified}),
at least for the strongly bound impurities.
To do this, one has to differentiate the expression for $\varkappa^{(W)} (\EF)$ with respect to
the impurity perturbation $\VL$.
Alternatively, we can reasonably assume that this maximum occurs when the Fermi energy $\EF$ lies
at the resonance half-width from the energy $\Er$. In this case,
the two terms in the denominator of  Eq.~(\ref{eqn:sigma-ata-lifshitz-simplified}) are approximately equal.
Using this property and taking the derivative over $\varepsilon$, we arrive the following rather simple estimate:
\begin{equation}
\label{eqn:kappa-m-1}
\varkappa_{\mathrm m}^{(W)} (\EF)
\approx
\frac{2}{\pi} \beta \ln \left(
\dfrac{W_0}{|\EF|} \right).
\end{equation}
One can see that Eq.~(\ref{eqn:kappa-m-1}) does not depend on the impurity perturbation $\VL$.
This expression is plotted in Fig.~\ref{fig:fano-kappa-efermi} as the
upper dotted-dashed gray line.
Taken with the opposite sign, it also provides a good estimate for the lower bound of the gain for the same
$\Delta \alpha = 0$ case. It is shown as the lower dotted-dashed gray line.

To derive a similar estimate for the term $\varkappa^{(\alpha)} (\EF)$ defined by Eq.~(\ref{varkappa-alpha}),
one has to use the Fano-Anderson model  with the impurity perturbation given by Eq.~(\ref{eqn:v-effective-alpha}).
Using the same simplifications as previously, we arrive at the following expression:
\begin{equation}
\label{eqn:kappa-m-alpha}
\varkappa_{\mathrm m}^{(\alpha)} (\EF)
\approx
-  \frac{W_0^2}{\pi \thyb^2} \frac{\Delta \alpha}{|\EF|}.
\end{equation}
Evidently, this term is nonzero only for $\Delta \alpha \neq 0$.
In this case the sum $\varkappa_{\mathrm m}^{(W)} (\EF)  +\varkappa_{\mathrm m}^{(\alpha)} (\EF)$ of the two contributions,
Eqs.~(\ref{eqn:kappa-m-1}) and (\ref{eqn:kappa-m-alpha}),
describes the upper bound of the gain for $\Delta \alpha = -\SI{2.5}{eV} $.
The same sum but taken with the negative sign, $-[\varkappa_{\mathrm m}^{(W)}(\EF) +\varkappa_{\mathrm m}^{(\alpha)} (\EF)]$,
provides the lower bound of the gain at particular value of $\EF$.
Both limits are shown in Fig.~\ref{fig:fano-kappa-efermi} by the double-dotted-dashed gray lines.

So far, we have analyzed the case of the strain-independent Fermi energy,
$|\EF^\varepsilon| = \mbox{const}$,
that for the considered concentration of impurities
corresponds to the constant Drude weight.
Now, we turn to the case of a sample with fixed number of carriers.

\subsection{Case of fixed number of carriers}

For the isolated sample with fixed number of carriers, one has to take into account the small
drift of the Fermi energy, $|\EF^\varepsilon| $, described by Eq.~(\ref{eqn:efermi-epsilon}).
As discussed in Sec.~\ref{sec:deformation-potential},  $1\%$ strain
results in the decrease of the interval $|\EF^\varepsilon|$ by only $\SI{3.5}{meV}$.
Note that for the impurity concentrations considered in this work, one can neglect
deviations of  the DOS from the clean graphene case [Eq.~(\ref{DOS-atom-cell}) caused by the resonance impurities].
Thus, we can use Eq.~(\ref{eqn:efermi-epsilon}) to obtain a value of the Fermi energy $\EF^\varepsilon$ in the strained sample.

Nevertheless, the obtained shift of the resonance energy is $\Er^\varepsilon - \Er \sim \SI{10}{meV}$
for $\varepsilon = 1\%$ and $\Er = -\SI{0.35}{eV}$.
Although this is three times larger than the corresponding shift of the Fermi energy, it is worth to
take into account the strain dependence of $\EF^\varepsilon$.

\begin{figure}[h]
  \centering
  \includegraphics[width=\linewidth]{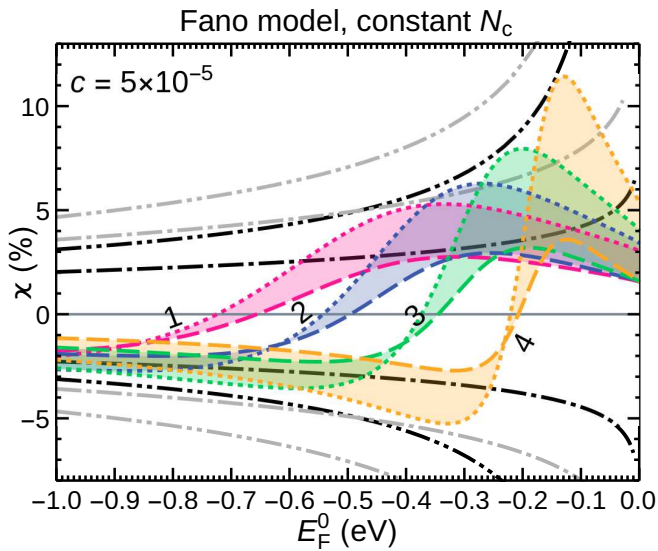}
\caption{The relative gain $\varkappa$ of the Drude width  per $1 \%$ of
strain as a function of the zero-strain Fermi energy $\EF^0 = W_0 \sqrt{\Nc}$ for the Fano-Anderson model.
The calculation is done assuming a constant strain independent carrier imbalance $\Nc$.
All parameters and notations are the same as in Fig.~\ref{fig:fano-kappa-efermi}.
The black dotted-dashed and double-dotted-dashed  lines are obtained on the base of analytical estimates
for  the upper and lower bounds of the gain function for the cases
$\alpha^\mathrm{imp} = \alpha$ and $\alpha^\mathrm{imp} =0$, respectively
(see the main text for the explanation).
The gray dotted-dashed and double-dotted-dashed gray lines are the same as in Fig.~\ref{fig:fano-kappa-efermi}.
}
\label{fig:kappa-fano-carriers}
\end{figure}

In Fig.~\ref{fig:kappa-fano-carriers} we plot the relative gain $\varkappa$ as a function of the zero-strain Fermi energy
$\EF^0 = W_0 \sqrt{\Nc}$. It is assumed that the number of carriers $\Nc$ is fixed, so that the value $\EF^\varepsilon$
changes under the strain.
Accordingly, the Drude width change is determined by both the resonance energy shift and the strain dependence of
the Fermi energy. We chose the same parameters and notations as in Fig.~\ref{fig:fano-kappa-efermi}.

We find that the relative gain depends on the Fermi energy similarly to the previous section,
but the magnitude of the effect is noticeably smaller.
While the results without accounting for the Fermi energy shift (see Fig.~\ref{fig:fano-kappa-efermi}) feature the values
of the relative gain up to $15 \%$ per $1\%$ of strain, in Fig.~\ref{fig:kappa-fano-carriers} we can see only a
$8 \%$ maximum increase for the same parameters. It is still large in comparison to the Born approximation,
which yields $\sim 3 \%$ regardless of the impurity parameters.

Qualitatively, one can understand the obtained results by the fact that the Fermi energy shifts in
the same direction as the resonance, albeit with a different strain rate.
This shift partially cancels the increase in the Drude width.
To account for this change in the calculation of the maximum gain,
$\varkappa_{\mathrm m} (\EF)$, we have to include
an extra term:
\begin{equation}
\begin{split}
\varkappa_{\mathrm m}^{(\EF)} (\EF) & =
\frac{1}{\Gamma_0}\left. \left(\dfrac{\partial
\Gamma}{\partial \EF^\varepsilon}\dfrac{\partial
\EF^\varepsilon}{\partial \varepsilon}\right)\right|_{\varepsilon = 0} \\
& \approx
\frac{\beta}{\pi} \left( - \ln \left(
\dfrac{W_0}{|\EF|} \right) + \frac{1}{2}\right).
\label{eqn:kappa-m-ef}
\end{split}
\end{equation}
The sums  $\pm [ \varkappa_{\mathrm m}^{(W)} (\EF) +
\varkappa_{\mathrm m}^{(\EF)} (\EF) +  \varkappa_{\mathrm m}^{(\alpha)} (\EF)]$
are shown in Fig.~\ref{fig:kappa-fano-carriers} as the dot-dashed and
double dot-dashed black lines for the cases
$\alpha^\mathrm{imp} = \alpha$ and $\alpha^\mathrm{imp} =0$, respectively.
For comparison, we also reproduced the corresponding  dot-dashed and
double dot-dashed gray lines from Fig.~\ref{fig:fano-kappa-efermi}
that do not include the term (\ref{eqn:kappa-m-ef}).

\section{Conclusions}
\label{sec:concl}

Our investigation of the properties of the Drude width in uniaxially strained graphene was partly motivated by the experimental work \cite{Kuzmenko20172DMat}.
As pointed out in \cite{Kuzmenko20172DMat}, the strong effect of strain on optical absorption of graphene at terahertz and lower
frequencies may have important implications for graphene-based optoelectronic devices, e.g., photodetectors, touch screens, and microelectromechanical systems.
The observation that these properties can be controlled mechanically opens new possibilities for the future applications of graphene.

We have thoroughly investigated a possible contribution of the point defects in the observed strong strain
dependence of the Drude width. A comparison with  other mechanisms of scattering of charge carriers,
which can overshadow the described effects, was not performed.
Instead we focused on in-depth study of impurities that can be described by
the Lifshitz and the Fano-Anderson models  within the framework of the ATA and the CPA approximations.
These approximations allow  to take into account the resonant character of these impurities that
cannot be properly addressed in the weak-scattering regime analyzed in Ref.~\onlinecite{Kuzmenko20172DMat}.

Another important issue considered in this work is the underlying mechanism of
the strain influence on the electronic spectrum of graphene.
The first effect considered in the vast majority of the literature
focuses on strain effect on hopping integrals.
The tensile strain results in the increase of the lattice bond lengths,  so the corresponding hopping parameters decrease.
This in  turn causes the deformation of the electronic spectrum and
the effective bandwidth $W_\varepsilon$, given by Eq.~(\ref{eqn:w-epsilon}), decreases.
In the weak-scattering regime, the impurity scattering rate is merely proportional to
the DOS [see Eq.~(\ref{DOS-atom-cell})].
The mentioned above decrease of the effective bandwidth $W_\varepsilon$ causes a moderate monotonic
increase in the DOS and corresponding scattering rate.

Discussed in this work  effects  are associated with the resonant impurity perturbation.
We show that it may cause a drastic increase in the impurity scattering rate when the corresponding
resonance is located in the vicinity of the Fermi level.
For the hole doped-sample shown in Fig.~\ref{fig:1}, the
resonance is assumed to be on the left side from the Fermi level.
As the strain is applied, the  interaction  between the impurity state and host results
both in the increase of the DOS and  shift of the impurity resonance towards  the Dirac point and thus
closer to the Fermi level.  In the case when the Drude weight is assumed to remain constant, this shift of the
resonance energy produces  a significant increase in the Drude-peak width.

In the case when the carrier number in the sample is fixed, the Fermi level also shifts toward
the Dirac point. Nevertheless, the  increase in the impurity scattering rate
still occurs because the Fermi level goes to the Dirac point slower than the resonance energy
as the strain is applied.

Yet, as we discussed, there is a second effect caused by the on-site deformation potential.
When this potential is the same both on the host and impurity sites, the whole picture
described above remains valid except  that the Dirac point, the Fermi level,
and the resonance energy are synchronously shifted by the same magnitude.
On the other hand, if we assume that the deformation potential changes the on-site energies of
the host atoms, while the energies of the the adatoms do not change, the resonant impurity scattering
rate is enhanced even more strongly under strain.

The effect of the strain on the impurity scattering rate is characterized by the dimensionless gain parameter
(\ref{eqn:relative-gain}), or by its discrete analog (\ref{eqn:relative-gain-1}), which is convenient to use
in numerical calculations when the dependence of the scattering rate on strain is linear.
Our main results that describe the gain for 1\% strain can be summarized as follows.

\begin{enumerate}
\item In the weak scattering limit we obtain that the gain is around 3\%, which is  larger than the
estimate done in Ref.~\onlinecite{Kuzmenko20172DMat}. The reason of the discrepancy is explained in Sec.~\ref{sec:Born}.

\item For the fixed Drude weight, $|\EF^\varepsilon| = |\cEF^\varepsilon - \cED^\varepsilon| = \mbox{const}$,
the maximal gain  is 8\%  in the case when the potential energy on impurity sites changes in concert with host sites.
The gain reaches 15\% when the potential on adatoms in the Fano-Anderson model does not change at all with the strain.

\item As the absolute value of the Fermi energy increases, the gain steeply diminishes
and  reaches zero when  $|\EF^\varepsilon |$ is approximately equal to the resonance energy.
Further increase of the Fermi energy results in the negative values of the gain
and then one reaches the minimal negative gain, i.e. the maximum decrease of the Drude weight.
The predicted characteristic nonmonotonic behavior of the gain as the function of the Fermi energy can be used to identify the contribution of the resonant impurities in the Drude width in the experiments with controllable carrier density \cite{Yu2016PRB}.

\item For the fixed carrier number when the difference $|\EF^\varepsilon |$
becomes strain dependent, the value of the maximal gain diminishes to 8\%.
\end{enumerate}

We emphasize that these values of the gain can be achieved when the impurity resonance
is located in the vicinity of the Fermi level that in the experiment \cite{Kuzmenko20172DMat} has
the specific value $\EF = \cEF - \cED \approx -\SI{0.23}{eV}$.
To conclude we note that even if the resonant impurities are not responsible for
the effects observed in Ref.~\onlinecite{Kuzmenko20172DMat}, this work allows one to speculate
that the resonant impurities added on graphene's sheet should allow one to control the corresponding
electronic properties of graphene.  This provides  guidelines  for functionalizing  graphene samples
in a way that would permit to modulate efficiently  the Drude-peak width by the applied strain.

\begin{acknowledgments}

We are indebted to A. Kuzmenko for bringing to our attention
his experimental results and for illuminating  discussions.
S.G.Sh. thanks members of Centre for Advanced 2D Materials, NUS for hospitality.
Y.V.S., S.G.Sh. and V.M.L. acknowledge the support
by the National Academy of Sciences of Ukraine (project KPKVK 6541230).
S.G.Sh. also acknowledges the support of EC for the
HORIZON 2020 RISE “CoExAN” Project (Project No. GA644076).

\end{acknowledgments}


\begin{thebibliography}{99}
\bibitem{foot} The corresponding relationship between the carrier (imbalance) density $n_\varepsilon$
and Fermi energy of noninteracting Dirac fermions
in uniaxially strained clean graphene reads
$
(\cEF^\varepsilon - \cED^\varepsilon)^2 \, \mbox{sgn} (\cEF^\varepsilon - \cED^\varepsilon)/(\pi \hbar^2 v_\parallel v_\perp )
= n_\varepsilon.
$
Using that $n_\varepsilon = n_0 S_0/S_\varepsilon \simeq n_0 (1 - \varepsilon)$, where $n_0$ is carrier imbalance
for unstrained graphene, one also obtains Eq.~(\ref{eqn:efermi-epsilon}).
\end{thebibliography}
\end{document}